\documentclass[11pt]{article}
\usepackage{fixltx2e} 
\usepackage{cmap} 
\usepackage[T1]{fontenc}
\usepackage[utf8]{inputenc}
\usepackage{palatino}
\usepackage{multirow}
\usepackage{ifthen}
\usepackage{amsmath,amssymb}
\usepackage{longtable,ltcaption,array}
\newlength{\DUtablewidth} 

\makeatletter
\usepackage{fullpage}
\usepackage{moresize}
\usepackage{graphicx}
\renewcommand{\ttfamily}{\fontfamily\ttdefault\selectfont\scriptsize}

\makeatother
\makeatletter

\RequirePackage{color}

\definecolor{DUcolor0}{rgb}{1.00,0.00,0.00}

\expandafter\providecommand\csname DUrolec1\endcsname[1]{\textit{#1}}

\expandafter\providecommand\csname DUroles2\endcsname[1]{\textit{#1}}

\expandafter\providecommand\csname DUroles1\endcsname[1]{\textit{#1}}

\makeatother

\providecommand{\DUadmonition}[2][class-arg]{%
  \ifcsname DUadmonition#1\endcsname%
    \csname DUadmonition#1\endcsname{#2}%
  \else
    \begin{center}
      \fbox{\parbox{0.9\textwidth}{#2}}
    \end{center}
  \fi
}

\providecommand*{\DUrole}[2]{%
  \ifcsname DUrole#1\endcsname%
    \csname DUrole#1\endcsname{#2}%
  \else
    \ifcsname docutilsrole#1\endcsname%
      \csname docutilsrole#1\endcsname{#2}%
    \else%
      #2%
    \fi%
  \fi%
}

\providecommand*{\DUtitle}[2][class-arg]{%
  \ifcsname DUtitle#1\endcsname%
    \csname DUtitle#1\endcsname{#2}%
  \else
    \smallskip\noindent\textbf{#2}\smallskip%
  \fi
}

\providecommand*{\DUtransition}[1][class-arg]{%
  \hspace*{\fill}\hrulefill\hspace*{\fill}
  \vskip 0.5\baselineskip
}

\ifthenelse{\isundefined{\hypersetup}}{
  \usepackage[colorlinks=true,linkcolor=blue,urlcolor=blue]{hyperref}
  \urlstyle{same} 
}{}
\hypersetup{
  pdftitle={Rust for functional programmers},
  pdfauthor={Raphael ‘kena’ Poss}
}

\title{\phantomsection%
  Rust for functional programmers%
  \label{rust-for-functional-programmers}}
\author{Raphael ‘kena’ Poss}
\date{July 2014}

\begin{document}
\maketitle

This post follows up on \href{http://blog.ezyang.com/2010/10/ocaml-for-haskellers/}{OCaml for Haskellers} from
Edward Z. Yang (2010) and my own \href{http://science.raphael.poss.name/haskell-for-ocaml-programmers.html}{Haskell for OCaml programmers}
from earlier this year.

\phantomsection\label{contents}
\pdfbookmark[1]{Contents}{contents}
\setcounter{tocdepth}{1}
\tableofcontents

\DUtransition

\DUadmonition[note]{
\DUtitle[note]{Note}

The latest version of this document can be found online at
\url{http://science.raphael.poss.name/rust-for-functional-programmers.html}.
Alternate formats:
\href{http://science.raphael.poss.name/rust-for-functional-programmers.txt}{Source},
\href{http://science.raphael.poss.name/rust-for-functional-programmers.pdf}{PDF}.
}

\section{Prologue%
  \label{prologue}%
}

Rust for C programmers != Rust for functional programmers.

For experienced C programmers with little experience in anything else,
Rust presents many new features and strange constructs that look and feel
relatively arcane.  Even to C++, Java or C\# programmers, Rust looks
foreign: its has objects but not classes, it has “structured enums”, a
strange “match” statement with no equivalent in the C family, it
prevents from assigning twice to the same variable, and all manners of
strange rules that were unheard of in the C family.

Why is that?

Although no current Rust manual dares putting it so bluntly, \textbf{Rust is
a functional language}, inspired from recent advances in
programming language design.

The problem faced by Rust advocates is that “functional programming,”
both as a term and as a discipline, has developed many stigmas over
the last 30 years: functional programs are “unreadable”, they are
relatively slow compared to C, they require heavyweight machinery
during execution (at least a garbage collector, often many functional
wrappers around system I/O), they are difficult to learn, they use a strange syntax, etc.

Trying to describe Rust as a functional language to C programmers, its
primary audience, would be a tough sell indeed. This is why the
official Rust manuals and tutorials duly and properly explain how to
use Rust for low-level tasks, and care to explain step-by-step the Rust
equivalents to many C programming patterns, without refering to other more modern languages.

\emph{This is all and well, but what if you already know about functional programming?}

For experienced users of Haskell, OCaml, etc., yet another
detailed manual that presents the basics of programming with
a functional flavor are a bore to read. Tailored to this audience, the
text below constitutes \textbf{a fast-track introduction to Rust}, from the
functional perspective.

\section{Why you should learn Rust%
  \label{why-you-should-learn-rust}%
}

For better or for worse, most hardware processors will continue to be
based on program counters, registers and addressable memory for the
foreseeable future. This is where we were in the 1970's, when C was
designed, and this is where we are still today, and this is precisely
why C is still in prevalent use.
Specifically, the C \emph{abstract machine model} is the snuggest fit for
most hardware platforms, and it is therefore a good level of abstraction to
build low-level system software like interrupt service
routines, garbage collectors and virtual memory managers.

However, the “user interface” of the C language, in particular \emph{its
preprocessor and its type system}, have aged tremendously.  For anyone
who learned anything newer, \emph{they frankly suck}.

This is where Rust has been designed: \textbf{Rust keeps the C abstract
machine model but innovates on the language interface}.  Rust is
expressive, its type system makes system code safer, and its powerful
meta-programming facilities enable new ways to generate code
automatically.

Note however that Rust is not yet fully stable at the time of this writing: it
is still subject to change with little notice, and the official documentation
is not fully synchronized with the implementation.

\DUtransition

\clearpage

\section{Straightforward equivalences%
  \label{straightforward-equivalences}%
}

\setlength{\DUtablewidth}{\linewidth}
\begin{longtable*}[c]{p{0.214\DUtablewidth}p{0.121\DUtablewidth}p{0.168\DUtablewidth}p{0.412\DUtablewidth}}
\textbf{%
Syntax
} & \textbf{%
Type
} & \textbf{%
Ocaml
} & \textbf{%
Haskell
} \\
\endfirsthead
\textbf{%
Syntax
} & \textbf{%
Type
} & \textbf{%
Ocaml
} & \textbf{%
Haskell
} \\
\endhead
\multicolumn{4}{c}{\hfill ... continued on next page} \\
\endfoot
\endlastfoot

\texttt{()}
 & 
()
 & 
\texttt{()}
 & 
\texttt{()}
 \\

\texttt{true}
 & 
bool
 & 
\texttt{true}
 & 
\texttt{True}
 \\

\texttt{false}
 & 
bool
 & 
\texttt{false}
 & 
\texttt{False}
 \\

\texttt{123}
 & 
(integer)
 &  & 
\texttt{123 \{- :: Num a => a \}}
 \\

\texttt{0x123}
 & 
(integer)
 &  & 
\texttt{0x123 \{- :: Num a => a \}}
 \\

\texttt{12.3}
 & 
(float)
 &  & 
\texttt{12.3 \{- :: Fractional a => a \}}
 \\

\texttt{'a'}
 & 
char
 &  & 
\texttt{'a'}
 \\

\texttt{\textquotedbl{}abc\textquotedbl{}}
 & 
str
 & 
\texttt{\textquotedbl{}abc\textquotedbl{}}
 & 
\texttt{\textquotedbl{}abc\textquotedbl{}}
 \\

\texttt{b'a'}
 & 
u8
 & 
\texttt{'a'}
 & 
\texttt{toEnum\$fromEnum\$'a'::Word8}
 \\

\texttt{123i}
 & 
int
 & 
\texttt{123}
 & 
\texttt{123 :: Int}
 \\

\texttt{123i32}
 & 
i32
 & 
\texttt{123l}
 & 
\texttt{123 :: Int32}
 \\

\texttt{123i64}
 & 
i64
 & 
\texttt{123L}
 & 
\texttt{123 :: Int64}
 \\

\texttt{123u}
 & 
uint
 &  & 
\texttt{123 :: Word}
 \\
\end{longtable*}

Like in Haskell, Rust number literals without a suffix do not have a
predefined type.  Their actual type is inferred from the context. Rust
character literals can represent any Unicode scalar value, in contrast
to OCaml's character which can only encode Latin-1 characters. Rust's
other literal forms are presented in \hyperref[literals]{Literals} below.

Primitive types:

\setlength{\DUtablewidth}{\linewidth}
\begin{longtable*}[c]{p{0.110\DUtablewidth}p{0.098\DUtablewidth}p{0.086\DUtablewidth}p{0.575\DUtablewidth}}
\textbf{%
Rust
} & \textbf{%
Haskell
} & \textbf{%
OCaml
} & \textbf{%
Description
} \\
\endfirsthead
\textbf{%
Rust
} & \textbf{%
Haskell
} & \textbf{%
OCaml
} & \textbf{%
Description
} \\
\endhead
\multicolumn{4}{c}{\hfill ... continued on next page} \\
\endfoot
\endlastfoot

()
 & 
()
 & 
unit
 & 
Unit type
 \\

bool
 & 
Bool
 & 
bool
 & 
Boolean type
 \\

int
 & 
Int
 & 
int
 & 
Signed integer, machine-dependent width
 \\

uint
 & 
Word
 &  & 
Unsigned integer, machine-dependent width
 \\

i8
 & 
Int8
 &  & 
8 bit wide signed integer, two's complement
 \\

i16
 & 
Int16
 &  & 
16 bit wide signed integer, two's complement
 \\

i32
 & 
Int32
 & 
int32
 & 
32 bit wide signed integer, two's complement
 \\

i64
 & 
Int64
 & 
int64
 & 
64 bit wide signed integer, two's complement
 \\

u8
 & 
Word8
 & 
char
 & 
8 bit wide unsigned integer
 \\

u16
 & 
Word16
 &  & 
16 bit wide unsigned integer
 \\

u32
 & 
Word32
 &  & 
32 bit wide unsigned integer
 \\

u64
 & 
Word64
 &  & 
64 bit wide unsigned integer
 \\

f32
 & 
Float
 &  & 
32-bit IEEE 754 binary floating-point
 \\

f64
 & 
Double
 & 
float
 & 
64-bit IEEE 754 binary floating-point
 \\

char
 & 
Char
 &  & 
Unicode scalar value (non-surrogate code points)
 \\

str
 &  &  & 
UTF-8 encoded character string.
 \\
\end{longtable*}

The type \texttt{str} in Rust is special: it is \emph{primitive}, so that the
compiler can optimize certain string operations; but it is not \emph{first
class}, so it is not possible to define variables of type \texttt{str} or
pass \texttt{str} values directly to functions. To use Rust strings in
programs, one should use string references, as described
later below.

Operators equivalences:
\begin{quote}{\ttfamily \raggedright \noindent
==~~!=~~<~~>~~<=~~>=~~\&\&~||~~~//~Rust\\
=~~~<>~~<~~>~~<=~~>=~~\&\&~||~~~(*~OCaml~*)\\
==~~/=~~<~~>~~<=~~>=~~\&\&~||~~~\{-~Haskell~-\}\\
~\\
+~~+~~+~~+~~-~~-~~*~~*~~/~~~~~/~~\%~~~~~!~~~~//~Rust\\
+~~+.~@~~\textasciicircum{}~~-~~-.~*~~*.~/~~~~~/.~mod~~~not~~(*~OCaml~*)\\
+~~+~~++~++~-~~-~~*~~*~~`div`~/~~`mod`~not~~\{-~Haskell~-\}\\
~\\
\&~~~~|~~~\textasciicircum{}~~~~<{}<~~~~~>{}>~~~~~!~~~~~~~~~~//~Rust\\
land~lor~lxor~{[}la{]}sl~{[}la{]}sr~lnot~~~~~~~(*~OCaml~*)\\
.\&.~~.|.~xor~~shiftL~shiftR~complement~\{-~Haskell~-\}
}
\end{quote}

Note that Rust uses \texttt{!} both for boolean negation and for the unary
bitwise NOT operator on integers. The unary \texttt{\textasciitilde{}} has a different
meaning in Rust than in C, detailed later below.

Compound expressions:

\setlength{\DUtablewidth}{\linewidth}
\begin{longtable*}[c]{p{0.166\DUtablewidth}p{0.193\DUtablewidth}p{0.166\DUtablewidth}p{0.425\DUtablewidth}}
\textbf{%
Rust
} & \textbf{%
OCaml
} & \textbf{%
Haskell
} & \textbf{%
Name
} \\
\endfirsthead
\textbf{%
Rust
} & \textbf{%
OCaml
} & \textbf{%
Haskell
} & \textbf{%
Name
} \\
\endhead
\multicolumn{4}{c}{\hfill ... continued on next page} \\
\endfoot
\endlastfoot

\texttt{R\{a:10, b:20\}}
 & 
\texttt{\{a=10; b=20\}}
 & 
\texttt{R\{a=10, b=20\}}
 & 
Record expression
 \\

\texttt{R\{a:30, ..z\}}
 & 
\texttt{\{z with a=30\}}
 & 
\texttt{z\{a=30\}}
 & 
Record with functional update
 \\

\texttt{(x,y,z)}
 & 
\texttt{(x,y,z)}
 & 
\texttt{(x,y,z)}
 & 
Tuple expression
 \\

\texttt{x.f}
 & 
\texttt{x.f}
 & 
\texttt{f x}
 & 
Field expression
 \\

\texttt{{[}x,y,z{]}}
 & 
\texttt{{[}|x;y;z|{]}}
 &  & 
Array expression, fixed size
 \\

\texttt{{[}x, ..10{]}}
 &  &  & 
Array expression, fixed repeats of first value
 \\

\texttt{x{[}10{]}}
 & 
\texttt{x.(10)}
 & 
\texttt{x!10}
 & 
Index expression (vectors/arrays)
 \\

\texttt{x{[}10{]}}
 & 
\texttt{x.{[}10{]}}
 & 
\texttt{x!!10}
 & 
Index expression (strings)
 \\

\texttt{\{x;y;z\}}
 & 
\texttt{begin x;y;z end}
 &  & 
Block expression
 \\

\texttt{\{x;y;\}}
 & 
\texttt{begin x;y;() end}
 &  & 
Block expression (ends with unit)
 \\
\end{longtable*}

Note that the value of a block expression is
the value of the last expression in the block, except when the block
ends with a semicolon, in which case its value is \texttt{()}.

Functions types and definitions:

\setlength{\DUtablewidth}{\linewidth}
\begin{longtable*}[c]{p{0.375\DUtablewidth}p{0.285\DUtablewidth}p{0.285\DUtablewidth}}

{\ttfamily \raggedright \noindent
\DUrole{c1}{//~Rust\\
//~f~:~|int,int|~->~int\\
}\DUrole{k}{fn}~\DUrole{n}{f}~\DUrole{p}{(}\DUrole{n}{x}\DUrole{o}{:}\DUrole{k}{int}\DUrole{p}{,}~\DUrole{n}{y}\DUrole{o}{:}\DUrole{k}{int}\DUrole{p}{)}~\DUrole{o}{->}~\DUrole{k}{int}~\DUrole{p}{\{}~\DUrole{n}{x}~\DUrole{o}{+}~\DUrole{n}{y}~\DUrole{p}{\};}~\\
~\\
\DUrole{c1}{//~fact~:~|int|~->~int\\
}\DUrole{k}{fn}~\DUrole{n}{fact}~\DUrole{p}{(}\DUrole{n}{n}\DUrole{o}{:}\DUrole{k}{int}\DUrole{p}{)}~\DUrole{o}{->}~\DUrole{k}{int}~\DUrole{p}{\{}~\\
~~~\DUrole{k}{if}~\DUrole{n}{n}~\DUrole{o}{==}~\DUrole{m}{1}\DUrole{k}{}~\DUrole{p}{\{}~\DUrole{m}{1}\DUrole{k}{}~\DUrole{p}{\}}~\\
~~~\DUrole{k}{else}~\DUrole{p}{\{}~\DUrole{n}{n}~\DUrole{o}{*}~\DUrole{n}{fact}\DUrole{p}{(}\DUrole{n}{n}\DUrole{o}{-}\DUrole{m}{1}\DUrole{k}{}\DUrole{p}{)}~\DUrole{p}{\}}~\\
\DUrole{p}{\}}
}
 & 
{\ttfamily \raggedright \noindent
\DUrole{c}{(*~OCaml~*)}~\\
\DUrole{c}{(*~val~f~:~int~*~int~->~int~*)}~\\
\DUrole{k}{let}~\DUrole{n}{f}~\DUrole{o}{(}\DUrole{n}{x}\DUrole{o}{,}~\DUrole{n}{y}\DUrole{o}{)}~\DUrole{o}{=}~\DUrole{n}{x}~\DUrole{o}{+}~\DUrole{n}{y}~\\
~\\
\DUrole{c}{(*~val~fact~:~int~->~int~*)}~\\
\DUrole{k}{let}~\DUrole{k}{rec}~\DUrole{n}{fact}~\DUrole{n}{n}~\DUrole{o}{=}~\\
~~~~\DUrole{k}{if}~\DUrole{n}{n}~\DUrole{o}{=}~\DUrole{mi}{1}~\DUrole{k}{then}~\DUrole{mi}{1}~\\
~~~~\DUrole{k}{else}~\DUrole{n}{n}~\DUrole{o}{*}~\DUrole{n}{fact}~\DUrole{o}{(}\DUrole{n}{n}\DUrole{o}{-}\DUrole{mi}{1}\DUrole{o}{)}
}
 & 
{\ttfamily \raggedright \noindent
\DUrole{cm}{\{-~Haskell~-\}}~\\
\DUrole{nf}{f}~\DUrole{ow}{::}~\DUrole{p}{(}\DUrole{kt}{Int}\DUrole{p}{,}~\DUrole{kt}{Int}\DUrole{p}{)}~\DUrole{ow}{->}~\DUrole{kt}{Int}~\\
\DUrole{nf}{f}~\DUrole{p}{(}\DUrole{n}{x}\DUrole{p}{,}~\DUrole{n}{y}\DUrole{p}{)}~\DUrole{ow}{=}~\DUrole{n}{x}~\DUrole{o}{+}~\DUrole{n}{y}~\\
~\\
\DUrole{nf}{fact}~\DUrole{ow}{::}~\DUrole{kt}{Int}~\DUrole{ow}{->}~\DUrole{kt}{Int}~\\
\DUrole{nf}{fact}~\DUrole{n}{n}~\DUrole{ow}{=}~\\
~~~~\DUrole{kr}{if}~\DUrole{n}{n}~\DUrole{ow}{=}~\DUrole{mi}{1}~\DUrole{kr}{then}~\DUrole{mi}{1}~\\
~~~~\DUrole{kr}{else}~\DUrole{n}{n}~\DUrole{o}{*}~\DUrole{n}{fact}\DUrole{p}{(}\DUrole{n}{n}\DUrole{o}{-}\DUrole{mi}{1}\DUrole{p}{)}
}
 \\
\end{longtable*}

Pattern match and guards:

\setlength{\DUtablewidth}{\linewidth}
\begin{longtable*}[c]{p{0.315\DUtablewidth}p{0.315\DUtablewidth}p{0.315\DUtablewidth}}

{\ttfamily \raggedright \noindent
\DUrole{c1}{//~Rust\\
}\DUrole{k}{match}~\DUrole{n}{e}~\DUrole{p}{\{}~\\
~~\DUrole{m}{0}\DUrole{k}{}~~~~~~~~~~~\DUrole{o}{=>}~\DUrole{m}{1}\DUrole{k}{}\DUrole{p}{,}~\\
~~\DUrole{n}{t}~\DUrole{o}{@}~\DUrole{m}{2}\DUrole{k}{}~~~~~~~\DUrole{o}{=>}~\DUrole{n}{t}~\DUrole{o}{+}~\DUrole{m}{1}\DUrole{k}{}\DUrole{p}{,}~\\
~~\DUrole{n}{n}~\DUrole{k}{if}~\DUrole{n}{n}~\DUrole{o}{<}~\DUrole{m}{10}\DUrole{k}{}~\DUrole{o}{=>}~\DUrole{m}{3}\DUrole{k}{}\DUrole{p}{,}~\\
~~\DUrole{n}{\_}~~~~~~~~~~~\DUrole{o}{=>}~\DUrole{m}{4}\DUrole{k}{}~\\
\DUrole{p}{\}}
}
 & 
{\ttfamily \raggedright \noindent
\DUrole{c}{(*~OCaml~*)}~\\
\DUrole{k}{match}~\DUrole{n}{e}~\DUrole{k}{with}~\\
~\DUrole{o}{|}~\DUrole{mi}{0}~~~~~~~~~~~~~\DUrole{o}{->}~\DUrole{mi}{1}~\\
~\DUrole{o}{|}~\DUrole{mi}{2}~\DUrole{k}{as}~\DUrole{n}{t}~~~~~~~~\DUrole{o}{->}~\DUrole{n}{t}~\DUrole{o}{+}~\DUrole{mi}{1}~\\
~\DUrole{o}{|}~\DUrole{n}{n}~\DUrole{k}{when}~\DUrole{n}{n}~\DUrole{o}{<}~\DUrole{mi}{10}~\DUrole{o}{->}~\DUrole{mi}{3}~\\
~\DUrole{o}{|}~\DUrole{o}{\_}~~~~~~~~~~~~~\DUrole{o}{->}~\DUrole{mi}{4}
}
 & 
{\ttfamily \raggedright \noindent
\DUrole{cm}{\{-~Haskell~-\}}~\\
\DUrole{kr}{case}~\DUrole{n}{e}~\DUrole{kr}{of}~\\
~~\DUrole{mi}{0}~~~~~~~~~~\DUrole{ow}{->}~\DUrole{mi}{1}~\\
~~\DUrole{n}{t}~\DUrole{o}{@}~\DUrole{mi}{2}~~~~~~\DUrole{ow}{->}~\DUrole{n}{t}~\DUrole{o}{+}~\DUrole{mi}{1}~\\
~~\DUrole{n}{n}~\DUrole{o}{|}~\DUrole{n}{n}~\DUrole{o}{<}~\DUrole{mi}{10}~\DUrole{ow}{->}~\DUrole{mi}{3}~\\
~~\DUrole{kr}{\_}~~~~~~~~~~\DUrole{ow}{->}~\DUrole{mi}{4}
}
 \\
\end{longtable*}

Recursion with side effects:

\setlength{\DUtablewidth}{\linewidth}
\begin{longtable*}[c]{p{0.315\DUtablewidth}p{0.315\DUtablewidth}p{0.315\DUtablewidth}}

{\ttfamily \raggedright \noindent
\DUrole{c1}{//~Rust\\
}\DUrole{k}{fn}~\DUrole{n}{collatz}\DUrole{p}{(}\DUrole{n}{n}\DUrole{o}{:}\DUrole{k}{uint}\DUrole{p}{)}~\DUrole{p}{\{}~\\
~~\DUrole{k}{let}~\DUrole{n}{v}~\DUrole{o}{=}~\DUrole{k}{match}~\DUrole{n}{n}~\DUrole{o}{\%}~\DUrole{m}{2}\DUrole{k}{}~\DUrole{p}{\{}~\\
~~~~~\DUrole{m}{0}\DUrole{k}{}~\DUrole{o}{=>}~\DUrole{n}{n}~\DUrole{o}{/}~\DUrole{m}{2}\DUrole{k}{}\DUrole{p}{,}~\\
~~~~~\DUrole{n}{\_}~\DUrole{o}{=>}~\DUrole{m}{3}\DUrole{k}{}~\DUrole{o}{*}~\DUrole{n}{n}~\DUrole{o}{+}~\DUrole{m}{1}\DUrole{k}{}~\\
~~\DUrole{p}{\}}~\\
~~\DUrole{n}{println}\DUrole{o}{!}\DUrole{p}{(}\DUrole{s}{\textquotedbl{}\{\}\textquotedbl{}}\DUrole{p}{,}~\DUrole{n}{v}\DUrole{p}{);}~\\
~~\DUrole{k}{if}~\DUrole{n}{v}~\DUrole{o}{!=}~\DUrole{m}{1}\DUrole{k}{}~\DUrole{p}{\{}~\DUrole{n}{collatz}\DUrole{p}{(}\DUrole{n}{v}\DUrole{p}{)}~\DUrole{p}{\}}~\\
\DUrole{p}{\}}~\\
\DUrole{k}{fn}~\DUrole{n}{main}\DUrole{p}{()}~\DUrole{p}{\{}~\DUrole{n}{collatz}\DUrole{p}{(}\DUrole{m}{25}\DUrole{k}{}\DUrole{p}{)}~\DUrole{p}{\}}
}
 & 
{\ttfamily \raggedright \noindent
\DUrole{c}{(*~OCaml~*)}~\\
\DUrole{k}{let}~\DUrole{k}{rec}~\DUrole{n}{collatz}~\DUrole{n}{n}~\DUrole{o}{=}~\\
~~~~\DUrole{k}{let}~\DUrole{n}{v}~\DUrole{o}{=}~\DUrole{k}{match}~\DUrole{n}{n}~\DUrole{o}{\%}~\DUrole{mi}{2}~\DUrole{k}{with}~\\
~~~~~~~\DUrole{o}{|}~\DUrole{mi}{0}~\DUrole{o}{->}~\DUrole{n}{n}~\DUrole{o}{/}~\DUrole{mi}{2}~\\
~~~~~~~\DUrole{o}{|}~\DUrole{o}{\_}~\DUrole{o}{->}~\DUrole{mi}{3}~\DUrole{o}{*}~\DUrole{n}{n}~\DUrole{o}{+}~\DUrole{mi}{1}~\\
~~~~\DUrole{k}{in}~\\
~~~~\DUrole{nn}{Printf}\DUrole{p}{.}\DUrole{n}{printf}~\DUrole{s2}{\textquotedbl{}\%d}\DUrole{se}{\textbackslash{}n}\DUrole{s2}{\textquotedbl{}}~\DUrole{n}{v}\DUrole{o}{;}~\\
~~~~\DUrole{k}{if}~\DUrole{n}{v}~\DUrole{o}{<>}~\DUrole{mi}{1}~\DUrole{k}{then}~\DUrole{n}{collatz}~\DUrole{n}{v}~\\
~\\
\DUrole{k}{let}~\DUrole{o}{\_}~\DUrole{o}{=}~\DUrole{n}{collatz}~\DUrole{mi}{25}
}
 & 
{\ttfamily \raggedright \noindent
\DUrole{cm}{\{-~Haskell~-\}}~\\
\DUrole{nf}{collatz}~\DUrole{n}{n}~\DUrole{ow}{=}~\DUrole{kr}{do}~\\
~~~~\DUrole{kr}{let}~\DUrole{n}{v}~\DUrole{ow}{=}~\DUrole{kr}{case}~\DUrole{n}{n}~\DUrole{p}{`}\DUrole{n}{mod}\DUrole{p}{`}~\DUrole{mi}{2}~\DUrole{kr}{of}~\\
~~~~~~~~~~~~\DUrole{mi}{0}~\DUrole{ow}{->}~\DUrole{n}{n}~\DUrole{p}{`}\DUrole{n}{div}\DUrole{p}{`}~\DUrole{mi}{2}~\\
~~~~~~~~~~~~\DUrole{kr}{\_}~\DUrole{ow}{->}~\DUrole{mi}{3}~\DUrole{o}{*}~\DUrole{n}{n}~\DUrole{o}{+}~\DUrole{mi}{1}~\\
~\\
~~~~\DUrole{n}{putStrLn}~\DUrole{o}{\$}~\DUrole{n}{show}~\DUrole{n}{v}~\\
~~~~\DUrole{n}{when}~\DUrole{p}{(}\DUrole{n}{v}~\DUrole{o}{/=}~\DUrole{mi}{1}\DUrole{p}{)}~\DUrole{o}{\$}~\DUrole{n}{collatz}~\DUrole{n}{v}~\\
~\\
\DUrole{nf}{main}~\DUrole{ow}{=}~\DUrole{n}{collatz}~\DUrole{mi}{25}
}
 \\
\end{longtable*}

Obviously, Rust uses strict (eager) evaluation and functions can
contain side effecting expressions, just like in OCaml.

Note that Rust does not (yet) guarantee tail call elimination,
although the underlying LLVM code generator is smart enough that it
should work for the function above. When in doubt, the following is equivalent:
\begin{quote}{\ttfamily \raggedright \noindent
\DUrole{k}{let}~\DUrole{k}{mut}~\DUrole{n}{n}~\DUrole{o}{=}~\DUrole{m}{25}\DUrole{k}{u}\DUrole{p}{;}~\\
\DUrole{k}{while}~\DUrole{n}{n}~\DUrole{o}{!=}~\DUrole{m}{1}\DUrole{k}{}~\DUrole{p}{\{}~\\
~~~\DUrole{n}{n}~\DUrole{o}{=}~\DUrole{k}{if}~\DUrole{n}{n}~\DUrole{o}{\%}~\DUrole{m}{2}\DUrole{k}{}~\DUrole{o}{==}~\DUrole{m}{0}\DUrole{k}{}~\DUrole{p}{\{}~\DUrole{n}{n}~\DUrole{o}{/}~\DUrole{m}{2}\DUrole{k}{}~\DUrole{p}{\}}~\\
~~~~~~~\DUrole{k}{else}~\DUrole{p}{\{}~\DUrole{m}{3}\DUrole{k}{}~\DUrole{o}{*}~\DUrole{n}{n}~\DUrole{o}{+}~\DUrole{m}{1}\DUrole{k}{}~\DUrole{p}{\}}~\\
~~~\DUrole{n}{println}\DUrole{p}{(}\DUrole{s}{\textquotedbl{}\{\}\textquotedbl{}}\DUrole{p}{,}~\DUrole{n}{n}\DUrole{p}{);}~\\
\DUrole{p}{\}}
}
\end{quote}

Record types, expressions and field access:

\setlength{\DUtablewidth}{\linewidth}
\begin{longtable*}[c]{p{0.315\DUtablewidth}p{0.315\DUtablewidth}p{0.315\DUtablewidth}}

{\ttfamily \raggedright \noindent
\DUrole{c1}{//~Rust\\
}\DUrole{k}{struct}~\DUrole{n}{Point}~\DUrole{p}{\{}~\\
~~\DUrole{n}{x}~\DUrole{o}{:}~\DUrole{k}{int}\DUrole{p}{,}~\\
~~\DUrole{n}{y}~\DUrole{o}{:}~\DUrole{k}{int}~\\
\DUrole{p}{\}}~\\
~\\
\DUrole{k}{let}~\DUrole{n}{v}~\DUrole{o}{=}~\DUrole{n}{Point}~\DUrole{p}{\{}\DUrole{n}{x}\DUrole{o}{:}\DUrole{m}{1}\DUrole{k}{}\DUrole{p}{,}~\DUrole{n}{y}\DUrole{o}{:}\DUrole{m}{2}\DUrole{k}{}\DUrole{p}{\};}~\\
\DUrole{k}{let}~\DUrole{n}{s}~\DUrole{o}{=}~\DUrole{n}{v}\DUrole{p}{.}\DUrole{n}{x}~\DUrole{o}{+}~\DUrole{n}{v}\DUrole{p}{.}\DUrole{n}{y}
}
 & 
{\ttfamily \raggedright \noindent
\DUrole{c}{(*~OCaml~*)}~\\
\DUrole{k}{type}~\DUrole{nc}{Point}~\DUrole{o}{=}~\DUrole{o}{\{}~\\
~~\DUrole{n}{x}~\DUrole{o}{:}~\DUrole{kt}{int}\DUrole{o}{;}~\\
~~\DUrole{n}{y}~\DUrole{o}{:}~\DUrole{kt}{int}~\\
\DUrole{o}{\}}~\\
~\\
\DUrole{k}{let}~\DUrole{n}{v}~\DUrole{o}{=}~\DUrole{o}{\{}~\DUrole{n}{x}~\DUrole{o}{=}~\DUrole{mi}{1}\DUrole{o}{;}~\DUrole{n}{y}~\DUrole{o}{=}~\DUrole{mi}{2}~\DUrole{o}{\};}~\\
\DUrole{k}{let}~\DUrole{n}{s}~\DUrole{o}{=}~\DUrole{n}{v}\DUrole{o}{.}\DUrole{n}{x}~\DUrole{o}{+}~\DUrole{n}{v}\DUrole{o}{.}\DUrole{n}{y}
}
 & 
{\ttfamily \raggedright \noindent
\DUrole{cm}{\{-~Haskell~-\}}~\\
\DUrole{kr}{data}~\DUrole{kt}{Point}~\DUrole{ow}{=}~\DUrole{kt}{Point}~\DUrole{p}{\{}~\\
~~~\DUrole{n}{x}~\DUrole{ow}{::}~\DUrole{kt}{Int}\DUrole{p}{,}~\\
~~~\DUrole{n}{y}~\DUrole{ow}{::}~\DUrole{kt}{Int}~\\
\DUrole{p}{\}}~\\
~\\
\DUrole{nf}{v}~\DUrole{ow}{=}~\DUrole{kt}{Point}~\DUrole{p}{\{}\DUrole{n}{x}~\DUrole{ow}{=}~\DUrole{mi}{1}\DUrole{p}{,}~\DUrole{n}{y}~\DUrole{ow}{=}~\DUrole{mi}{2}\DUrole{p}{\}}~\\
\DUrole{nf}{s}~\DUrole{ow}{=}~\DUrole{p}{(}\DUrole{n}{x}~\DUrole{n}{v}\DUrole{p}{)}~\DUrole{o}{+}~\DUrole{p}{(}\DUrole{n}{y}~\DUrole{n}{v}\DUrole{p}{)}
}
 \\
\end{longtable*}

Free type parameters (generic data and function types):

\setlength{\DUtablewidth}{\linewidth}
\begin{longtable*}[c]{p{0.315\DUtablewidth}p{0.315\DUtablewidth}p{0.315\DUtablewidth}}

{\ttfamily \raggedright \noindent
\DUrole{c1}{//~Rust\\
}\DUrole{k}{type}~\DUrole{n}{Pair}\DUrole{o}{<}\DUrole{n}{a}\DUrole{p}{,}\DUrole{n}{b}\DUrole{o}{>}~\DUrole{o}{=}~\DUrole{p}{(}\DUrole{n}{a}\DUrole{p}{,}\DUrole{n}{b}\DUrole{p}{)}~\\
~\\
\DUrole{c1}{//~id<t>~:~|t|~->~t\\
}\DUrole{k}{fn}~\DUrole{n}{id}\DUrole{o}{<}\DUrole{n}{t}\DUrole{o}{>}\DUrole{p}{(}\DUrole{n}{x}~\DUrole{o}{:}~\DUrole{n}{t}\DUrole{p}{)}~\DUrole{o}{->}~\DUrole{n}{t}~\DUrole{p}{\{}~\DUrole{n}{x}~\DUrole{p}{\}}
}
 & 
{\ttfamily \raggedright \noindent
\DUrole{c}{(*~OCaml~*)}~\\
\DUrole{k}{type}~\DUrole{o}{(}\DUrole{k}{'}\DUrole{n}{a}\DUrole{o}{,}~\DUrole{k}{'}\DUrole{n}{b}\DUrole{o}{)}~\DUrole{n}{pair}~\DUrole{o}{=}~\DUrole{k}{'}\DUrole{n}{a}~\DUrole{o}{*}~\DUrole{k}{'}\DUrole{n}{b}~\\
~\\
\DUrole{c}{(*~val~id~:~'t~->~'t~*)}~\\
\DUrole{k}{let}~\DUrole{n}{id}~\DUrole{n}{x}~\DUrole{o}{=}~\DUrole{n}{x}
}
 & 
{\ttfamily \raggedright \noindent
\DUrole{cm}{\{-~Haskell~-\}}~\\
\DUrole{kr}{type}~\DUrole{kt}{Pair}~\DUrole{n}{a}~\DUrole{n}{b}~\DUrole{ow}{=}~\DUrole{p}{(}\DUrole{n}{a}\DUrole{p}{,}~\DUrole{n}{b}\DUrole{p}{)}~\\
~\\
\DUrole{nf}{id}~\DUrole{ow}{::}~\DUrole{n}{t}~\DUrole{ow}{->}~\DUrole{n}{t}~\\
\DUrole{nf}{id}~\DUrole{n}{x}~\DUrole{ow}{=}~\DUrole{n}{x}
}
 \\
\end{longtable*}

Algebraic data types:

\setlength{\DUtablewidth}{\linewidth}
\begin{longtable*}[c]{p{0.315\DUtablewidth}p{0.315\DUtablewidth}p{0.315\DUtablewidth}}

{\ttfamily \raggedright \noindent
\DUrole{c1}{//~Rust\\
}\DUrole{k}{enum}~\DUrole{n}{Option}\DUrole{o}{<}\DUrole{n}{T}\DUrole{o}{>}~\DUrole{p}{\{}~\\
~~~\DUrole{n}{None}\DUrole{p}{,}~\\
~~~\DUrole{n}{Some}\DUrole{p}{(}\DUrole{n}{T}\DUrole{p}{)}~\\
\DUrole{p}{\}}~\\
\DUrole{c1}{//~x~:~Option<t>\\
}\DUrole{k}{match}~\DUrole{n}{x}~\DUrole{p}{\{}~\\
~~\DUrole{n}{None}~~~~\DUrole{o}{=>}~\DUrole{k}{false}\DUrole{p}{,}~\\
~~\DUrole{n}{Some}\DUrole{p}{(}\DUrole{n}{\_}\DUrole{p}{)}~\DUrole{o}{=>}~\DUrole{k}{true}~\\
\DUrole{p}{\}}
}
 & 
{\ttfamily \raggedright \noindent
\DUrole{c}{(*~OCaml~*)}~\\
\DUrole{k}{type}~\DUrole{k}{'}\DUrole{n}{t}~\DUrole{n}{option}~\DUrole{o}{=}~\\
~~~~\DUrole{nc}{None}~\\
~~\DUrole{o}{|}~\DUrole{nc}{Some}~\DUrole{k}{of}~\DUrole{k}{'}\DUrole{n}{t}~\\
~\\
\DUrole{c}{(*~x~:~t~option~*)}~\\
\DUrole{k}{match}~\DUrole{n}{x}~\DUrole{k}{with}~\\
~~\DUrole{o}{|}~\DUrole{nc}{None}~\DUrole{o}{->}~\DUrole{bp}{false}~\\
~~\DUrole{o}{|}~\DUrole{nc}{Some}~\DUrole{o}{\_}~\DUrole{o}{->}~\DUrole{bp}{true}
}
 & 
{\ttfamily \raggedright \noindent
\DUrole{cm}{\{-~Haskell~-\}}~\\
\DUrole{kr}{data}~\DUrole{kt}{Maybe}~\DUrole{n}{a}~\DUrole{ow}{=}~\\
~~~~\DUrole{kt}{Nothing}~\\
~~\DUrole{o}{|}~\DUrole{kt}{Just}~\DUrole{n}{a}~\\
~\\
\DUrole{cm}{\{-~x~:~Maybe~t~-\}}~\\
\DUrole{kr}{case}~\DUrole{n}{x}~\DUrole{kr}{of}~\\
~~~\DUrole{kt}{Nothing}~\DUrole{ow}{->}~\DUrole{kt}{False}~\\
~~~\DUrole{kt}{Just}~\DUrole{kr}{\_}~\DUrole{ow}{->}~\DUrole{kt}{True}
}
 \\
\end{longtable*}

Lambda expressions and higher-order functions:

\setlength{\DUtablewidth}{\linewidth}
\begin{longtable*}[c]{p{0.375\DUtablewidth}p{0.285\DUtablewidth}p{0.285\DUtablewidth}}

{\ttfamily \raggedright \noindent
\DUrole{c1}{//~Rust\\
//~||int,int|~->~int,~int|~->~int\\
}\DUrole{k}{fn}~\DUrole{n}{ff}\DUrole{p}{(}\DUrole{n}{f}\DUrole{o}{:|}\DUrole{k}{int}\DUrole{p}{,}\DUrole{k}{int}\DUrole{o}{|->}\DUrole{k}{int}\DUrole{p}{,}~\DUrole{n}{x}\DUrole{o}{:}\DUrole{k}{int}\DUrole{p}{)}~\DUrole{o}{->}~\DUrole{k}{int}~\\
\DUrole{p}{\{}~\DUrole{n}{f}~\DUrole{p}{(}\DUrole{n}{x}\DUrole{p}{,}~\DUrole{n}{x}\DUrole{p}{)}~\DUrole{p}{\}}~\\
~\\
\DUrole{c1}{//~m2~:~|int|~->~int\\
}\DUrole{k}{fn}~\DUrole{n}{m2}\DUrole{p}{(}\DUrole{n}{n}~\DUrole{o}{:}~\DUrole{k}{int}\DUrole{p}{)}~\DUrole{o}{->}~\DUrole{k}{int}~\\
\DUrole{p}{\{}~\DUrole{n}{ff}~\DUrole{p}{((}\DUrole{o}{|}\DUrole{n}{x}\DUrole{p}{,}\DUrole{n}{y}\DUrole{o}{|}~\DUrole{p}{\{}~\DUrole{n}{x}~\DUrole{o}{+}~\DUrole{n}{y}~\DUrole{p}{\}),}~\DUrole{n}{n}\DUrole{p}{)}~\DUrole{p}{\}}
}
 & 
{\ttfamily \raggedright \noindent
\DUrole{c}{(*~OCaml~*)}~\\
\DUrole{c}{(*~(int*int->b)*int~->~int~*)}~\\
\DUrole{k}{let}~\DUrole{n}{ff}~\DUrole{o}{(}\DUrole{n}{f}\DUrole{o}{,}~\DUrole{n}{x}\DUrole{o}{)}~\DUrole{o}{=}~\\
~~~~\DUrole{n}{f}~\DUrole{o}{(}\DUrole{n}{x}\DUrole{o}{,}~\DUrole{n}{x}\DUrole{o}{)}~\\
~\\
\DUrole{c}{(*~m2~:~int~->~int~*)}~\\
\DUrole{k}{let}~\DUrole{n}{m2}~\DUrole{n}{n}~\DUrole{o}{=}~\\
~~~~\DUrole{n}{ff}~\DUrole{o}{((}\DUrole{k}{fun}\DUrole{o}{(}\DUrole{n}{x}\DUrole{o}{,}\DUrole{n}{y}\DUrole{o}{)}~\DUrole{o}{->}~\DUrole{n}{x}~\DUrole{o}{+}~\DUrole{n}{y}\DUrole{o}{),}~\DUrole{n}{n}\DUrole{o}{)}
}
 & 
{\ttfamily \raggedright \noindent
\DUrole{cm}{\{-~Haskell~-\}}~\\
\DUrole{nf}{ff}~\DUrole{ow}{::}~\DUrole{p}{((}\DUrole{n}{int}\DUrole{p}{,}\DUrole{n}{int}\DUrole{p}{)}\DUrole{ow}{->}\DUrole{n}{int}\DUrole{p}{,}~\DUrole{n}{int}\DUrole{p}{)}~\DUrole{ow}{->}~\DUrole{n}{int}~\\
\DUrole{nf}{ff}~\DUrole{p}{(}\DUrole{n}{f}\DUrole{p}{,}~\DUrole{n}{x}\DUrole{p}{)}~\DUrole{ow}{=}~\\
~~~\DUrole{n}{f}~\DUrole{p}{(}\DUrole{n}{x}\DUrole{p}{,}~\DUrole{n}{x}\DUrole{p}{)}~\\
~\\
\DUrole{nf}{m2}~\DUrole{ow}{::}~\DUrole{kt}{Int}~\DUrole{ow}{->}~\DUrole{kt}{Int}~\\
\DUrole{nf}{m2}~\DUrole{n}{n}~\DUrole{ow}{=}~\\
~~~\DUrole{n}{ff}~\DUrole{p}{((}\DUrole{nf}{\textbackslash{}}\DUrole{p}{(}\DUrole{n}{x}\DUrole{p}{,}\DUrole{n}{y}\DUrole{p}{)}~\DUrole{ow}{->}~\DUrole{n}{x}~\DUrole{o}{+}~\DUrole{n}{y}\DUrole{p}{),}~\DUrole{n}{n}\DUrole{p}{)}
}
 \\
\end{longtable*}

\section{Traits: Rust's type classes%
  \label{traits-rust-s-type-classes}%
}

Rust's “traits” are analogous to Haskell's type classes.

The main difference with Haskell is that traits only intervene for
expressions with dot notation, ie. of the form \texttt{a.foo(b)}.

For C++/Java/C\#/OCaml programmers, however, traits should
not be confused with traditional object classes. They are
really type classes: it is possible to add traits to arbitrary
data types, including the primitive types!

An example:

\setlength{\DUtablewidth}{\linewidth}
\begin{longtable*}[c]{p{0.470\DUtablewidth}p{0.470\DUtablewidth}}

{\ttfamily \raggedright \noindent
\DUrole{c1}{//~Rust\\
}\DUrole{k}{trait}~\DUrole{n}{Testable}~\DUrole{p}{\{}~\\
~~~~\DUrole{k}{fn}~\DUrole{n}{test}\DUrole{p}{(}\DUrole{o}{\&}\DUrole{n}{self}\DUrole{p}{)}~\DUrole{o}{->}~\DUrole{n}{bool}~\\
\DUrole{p}{\}}~\\
~\\
\DUrole{k}{impl}~\DUrole{n}{Testable}~\DUrole{k}{for}~\DUrole{k}{int}~\DUrole{p}{\{}~\\
~~~~\DUrole{k}{fn}~\DUrole{n}{test}\DUrole{p}{(}\DUrole{o}{\&}\DUrole{n}{self}\DUrole{p}{)}~\DUrole{o}{->}~\DUrole{k}{int}~\DUrole{p}{\{}~\\
~~~~~~~\DUrole{k}{if}~\DUrole{o}{*}\DUrole{n}{self}~\DUrole{o}{==}~\DUrole{m}{0}\DUrole{k}{}~\DUrole{p}{\{}~\DUrole{k}{false}~\DUrole{p}{\}}~\\
~~~~~~~\DUrole{k}{else}~\DUrole{p}{\{}~\DUrole{k}{true}~\DUrole{p}{\}}~\\
~~~~\DUrole{p}{\}}~\\
\DUrole{p}{\}}~\\
~\\
\DUrole{k}{fn}~\DUrole{n}{hello}\DUrole{p}{(}\DUrole{n}{x}\DUrole{o}{:}\DUrole{k}{int}\DUrole{p}{)}~\DUrole{o}{->}~\DUrole{n}{bool}~\DUrole{p}{\{}~\\
~~~\DUrole{n}{x}\DUrole{p}{.}\DUrole{n}{test}\DUrole{p}{()}~\\
\DUrole{p}{\}}
}
 & 
{\ttfamily \raggedright \noindent
\DUrole{cm}{\{-~Haskell~-\}}~\\
\DUrole{kr}{class}~\DUrole{kt}{Testable}~\DUrole{n}{a}~\DUrole{kr}{where}~\\
~~~~\DUrole{n}{test}~\DUrole{ow}{::}~\DUrole{n}{a}~\DUrole{ow}{->}~\DUrole{kt}{Bool}~\\
~\\
~\\
\DUrole{kr}{instance}~\DUrole{kt}{Testable}~\DUrole{kt}{Int}~\DUrole{kr}{where}~\\
~~~~\DUrole{n}{test}~\DUrole{n}{n}~\DUrole{ow}{=}~\\
~~~~~~~\DUrole{kr}{if}~\DUrole{n}{n}~\DUrole{o}{==}~\DUrole{mi}{0}~\DUrole{kr}{then}~\DUrole{kt}{False}~\\
~~~~~~~\DUrole{kr}{else}~\DUrole{kt}{True}~\\
~\\
~\\
\DUrole{nf}{hello}~\DUrole{ow}{::}~\DUrole{kt}{Int}~\DUrole{ow}{->}~\DUrole{kt}{Bool}~\\
\DUrole{nf}{hello}~\DUrole{n}{x}~\DUrole{ow}{=}~\\
~~~~\DUrole{n}{test}~\DUrole{n}{x}
}
 \\
\end{longtable*}

In a trait method declaration, the identifier “\texttt{self}” denotes
the actual object on which the method is applied.

Like in Haskell, Rust traits can be used for operator overloading. For
example, if one defines a new sum type for Peano integers:

\setlength{\DUtablewidth}{\linewidth}
\begin{longtable*}[c]{p{0.470\DUtablewidth}p{0.470\DUtablewidth}}

{\ttfamily \raggedright \noindent
\DUrole{c1}{//~Rust\\
}\DUrole{k}{enum}~\DUrole{n}{Peano}~\DUrole{p}{\{}~\\
~~~\DUrole{n}{Zero}\DUrole{p}{,}~\\
~~~\DUrole{n}{Succ}\DUrole{p}{(}\DUrole{n}{Box}\DUrole{o}{<}\DUrole{n}{Peano}\DUrole{o}{>}\DUrole{p}{)}~\\
\DUrole{p}{\}}
}
 & 
{\ttfamily \raggedright \noindent
\DUrole{cm}{\{-~Haskell~-\}}~\\
\DUrole{kr}{data}~\DUrole{kt}{Peano}~\DUrole{ow}{=}~\\
~~~~\DUrole{kt}{Zero}~\\
~~\DUrole{o}{|}~\DUrole{kt}{Succ}~\DUrole{kt}{Peano}
}
 \\
\end{longtable*}

Then one can overload the comparison operator \texttt{==} between Peano
integers by instantiating the \texttt{PartialEq} class:

\setlength{\DUtablewidth}{\linewidth}
\begin{longtable*}[c]{p{0.470\DUtablewidth}p{0.470\DUtablewidth}}

{\ttfamily \raggedright \noindent
\DUrole{c1}{//~Rust\\
}\DUrole{k}{impl}~\DUrole{n}{PartialEq}~\DUrole{k}{for}~\DUrole{n}{Peano}~\DUrole{p}{\{}~\\
~~~~\DUrole{k}{fn}~\DUrole{n}{eq}\DUrole{p}{(}\DUrole{o}{\&}\DUrole{n}{self}\DUrole{p}{,}~\DUrole{n}{other}\DUrole{o}{:\&}\DUrole{n}{Peano}\DUrole{p}{)}~\DUrole{o}{->}~\DUrole{n}{bool}~\DUrole{p}{\{}~\\
~~~~~~\DUrole{k}{match}~\DUrole{p}{(}\DUrole{n}{self}\DUrole{p}{,}~\DUrole{n}{other}\DUrole{p}{)}~\DUrole{p}{\{}~\\
~~~~~~~~\DUrole{p}{(}\DUrole{o}{\&}\DUrole{n}{Zero}\DUrole{p}{,}~\DUrole{o}{\&}\DUrole{n}{Zero}\DUrole{p}{)}~~~~~~~~~~~~~~~\DUrole{o}{=>}~\DUrole{k}{true}\DUrole{p}{,}~\\
~~~~~~~~\DUrole{p}{(}\DUrole{o}{\&}\DUrole{n}{Succ}\DUrole{p}{(}\DUrole{k}{ref}~\DUrole{n}{a}\DUrole{p}{),}~\DUrole{o}{\&}\DUrole{n}{Succ}\DUrole{p}{(}\DUrole{k}{ref}~\DUrole{n}{b}\DUrole{p}{))}~\DUrole{o}{=>}~\DUrole{p}{(}\DUrole{n}{a}~\DUrole{o}{==}~\DUrole{n}{b}\DUrole{p}{),}~\\
~~~~~~~~\DUrole{p}{(}\DUrole{n}{\_}\DUrole{p}{,}~\DUrole{n}{\_}\DUrole{p}{)}~~~~~~~~~~~~~~~~~~~~~~~\DUrole{o}{=>}~\DUrole{k}{false}~\\
~~~~~~\DUrole{p}{\}}~\\
~~~~\DUrole{p}{\}}~\\
\DUrole{p}{\}}
}
 & 
{\ttfamily \raggedright \noindent
\DUrole{cm}{\{-~Haskell~-\}}~\\
\DUrole{kr}{instance}~\DUrole{kt}{Eq}~\DUrole{kt}{Peano}~\DUrole{kr}{where}~\\
~~~~\DUrole{p}{(}\DUrole{o}{==}\DUrole{p}{)}~\DUrole{n}{self}~\DUrole{n}{other}~\DUrole{ow}{=}~\\
~~~~~~~~\DUrole{kr}{case}~\DUrole{p}{(}\DUrole{n}{self}\DUrole{p}{,}~\DUrole{n}{other}\DUrole{p}{)}~\DUrole{kr}{of}~\\
~~~~~~~~~~\DUrole{o}{|}~\DUrole{p}{(}\DUrole{kt}{Zero}\DUrole{p}{,}~\DUrole{kt}{Zero}\DUrole{p}{)}~~~~~\DUrole{ow}{->}~\DUrole{kt}{True}~\\
~~~~~~~~~~\DUrole{o}{|}~\DUrole{p}{(}\DUrole{kt}{Succ}~\DUrole{n}{a}\DUrole{p}{,}~\DUrole{kt}{Succ}~\DUrole{n}{b}\DUrole{p}{)}~\DUrole{ow}{->}~\DUrole{p}{(}\DUrole{n}{a}~\DUrole{o}{==}~\DUrole{n}{b}\DUrole{p}{)}~\\
~~~~~~~~~~\DUrole{o}{|}~\DUrole{p}{(}\DUrole{kr}{\_}\DUrole{p}{,}~\DUrole{kr}{\_}\DUrole{p}{)}~~~~~~~~~~~\DUrole{ow}{->}~\DUrole{kt}{False}
}
 \\
\end{longtable*}

Also, like in Haskell, a trait can provide a default implementation
for a method, to be used when instances omit the specialization:

\setlength{\DUtablewidth}{\linewidth}
\begin{longtable*}[c]{p{0.470\DUtablewidth}p{0.470\DUtablewidth}}

{\ttfamily \raggedright \noindent
\DUrole{c1}{//~Rust\\
}\DUrole{k}{trait}~\DUrole{n}{PartialEq}~\DUrole{p}{\{}~\\
~~~\DUrole{k}{fn}~\DUrole{n}{eq}\DUrole{p}{(}\DUrole{o}{\&}\DUrole{n}{self}\DUrole{p}{,}~\DUrole{n}{other}\DUrole{o}{:\&}\DUrole{n}{Self}\DUrole{p}{)}~\DUrole{o}{->}~\DUrole{n}{bool}\DUrole{p}{;}~\\
~~~\DUrole{k}{fn}~\DUrole{n}{ne}\DUrole{p}{(}\DUrole{o}{\&}\DUrole{n}{self}\DUrole{p}{,}~\DUrole{n}{other}\DUrole{o}{:\&}\DUrole{n}{Self}\DUrole{p}{)}~\DUrole{o}{->}~\DUrole{n}{bool}~\\
~~~\DUrole{p}{\{}~\DUrole{o}{!}\DUrole{n}{self}\DUrole{p}{.}\DUrole{n}{eq}\DUrole{p}{(}\DUrole{n}{other}\DUrole{p}{)}~\DUrole{p}{\}}~\\
~\DUrole{p}{\}}
}
 & 
{\ttfamily \raggedright \noindent
\DUrole{cm}{\{-~Haskell~-\}}~\\
\DUrole{kr}{class}~\DUrole{kt}{Eq}~\DUrole{n}{a}~\DUrole{kr}{where}~\\
~~~\DUrole{p}{(}\DUrole{o}{==}\DUrole{p}{)}~\DUrole{kt}{:}~\DUrole{n}{a}~\DUrole{ow}{->}~\DUrole{n}{a}~\DUrole{ow}{->}~\DUrole{kt}{Bool}~\\
~~~\DUrole{p}{(}\DUrole{o}{!=}\DUrole{p}{)}~\DUrole{kt}{:}~\DUrole{n}{a}~\DUrole{ow}{->}~\DUrole{n}{a}~\DUrole{ow}{->}~\DUrole{kt}{Bool}~\\
~~~\DUrole{p}{(}\DUrole{o}{!=}\DUrole{p}{)}~\DUrole{n}{x}~\DUrole{n}{y}~\DUrole{ow}{=}~\DUrole{n}{not}~\DUrole{p}{(}\DUrole{n}{x}~\DUrole{o}{==}~\DUrole{n}{y}\DUrole{p}{)}
}
 \\
\end{longtable*}

In the method declarations inside a trait declaration, the identifier
“\texttt{Self}” refers to the actual type on which the trait applies.

Each overloadable operator in Rust has a corresponding
trait in the standard library:

\setlength{\DUtablewidth}{\linewidth}
\begin{longtable*}[c]{p{0.106\DUtablewidth}p{0.149\DUtablewidth}p{0.199\DUtablewidth}p{0.495\DUtablewidth}}
\textbf{%
Expression
} & \textbf{%
Expands to
} & \textbf{%
Trait
} & \textbf{%
Equivalent Haskell class/method
} \\
\endfirsthead
\textbf{%
Expression
} & \textbf{%
Expands to
} & \textbf{%
Trait
} & \textbf{%
Equivalent Haskell class/method
} \\
\endhead
\multicolumn{4}{c}{\hfill ... continued on next page} \\
\endfoot
\endlastfoot

\texttt{a == b}
 & 
\texttt{a.eq(b)}
 & 
std::cmp::PartialEq
 & 
\texttt{class PartialEq a where (==) : a -> a -> bool}
 \\

\texttt{a != b}
 & 
\texttt{a.ne(b)}
 & 
std::cmp::PartialEq
 & 
\texttt{class PartialEq a where (/=) : a -> a -> bool}
 \\

\texttt{a < ~b}
 & 
\texttt{a.lt(b)}
 & 
std::cmp::PartialOrd
 & 
\texttt{class PartialOrd a where (<) : a -> a -> bool}
 \\

\texttt{a > ~b}
 & 
\texttt{a.gt(b)}
 & 
std::cmp::PartialOrd
 & 
\texttt{class PartialOrd a where (>) : a -> a -> bool}
 \\

\texttt{a <= b}
 & 
\texttt{a.le(b)}
 & 
std::cmp::PartialOrd
 & 
\texttt{class PartialOrd a : Eq a where (<=) : a -> a -> bool}
 \\

\texttt{a >= b}
 & 
\texttt{a.ge(b)}
 & 
std::cmp::PartialOrd
 & 
\texttt{class PartialOrd a : Eq a where (>=) : a -> a -> bool}
 \\

\texttt{a + b}
 & 
\texttt{a.add(b)}
 & 
std::ops::Add<b,c>
 & 
\texttt{class Add a b c where (+) : a -> b -> c}
 \\

\texttt{a - b}
 & 
\texttt{a.sub(b)}
 & 
std::ops::Sub<b,c>
 & 
\texttt{class Sub a b c where (-) : a -> b -> c}
 \\

\texttt{a * b}
 & 
\texttt{a.mul(b)}
 & 
std::ops::Mul<b,c>
 & 
\texttt{class Mul a b c where (*) : a -> b -> c}
 \\

\texttt{a / b}
 & 
\texttt{a.div(b)}
 & 
std::ops::Div<b,c>
 & 
\texttt{class Div a b c where (/) : a -> b -> c}
 \\

\texttt{a \% b}
 & 
\texttt{a.rem(b)}
 & 
std::ops::Rem<b,c>
 & 
\texttt{class Rem a b c where (\%) : a -> b -> c}
 \\

\texttt{-a}
 & 
\texttt{a.neg()}
 & 
std::ops::Neg<c>
 & 
\texttt{class Neg a c where (-) : a -> c}
 \\

\texttt{!a}
 & 
\texttt{a.not()}
 & 
std::ops::Not<c>
 & 
\texttt{class Not a c where (!) : a -> c}
 \\

\texttt{*a}
 & 
\texttt{a.deref()}
 & 
std::ops::Deref<c>
 & 
\texttt{class Deref a c where (*) : a -> c}
 \\

\texttt{a \& b}
 & 
\texttt{a.bitand(b)}
 & 
std::ops::BitAnd<b,c>
 & 
\texttt{class BitAnd a b c where (\&) : a -> b -> c}
 \\

\texttt{a | b}
 & 
\texttt{a.bitor(b)}
 & 
std::ops::BitOr<b,c>
 & 
\texttt{class BitOr a b c where (|) : a -> b -> c}
 \\

\texttt{a \textasciicircum{} b}
 & 
\texttt{a.bitxor(b)}
 & 
std::ops::BitXor<b,c>
 & 
\texttt{class BitXor a b c where (\textasciicircum{}) : a -> b -> c}
 \\

\texttt{a <{}< b}
 & 
\texttt{a.shl(b)}
 & 
std::ops::Shl<b,c>
 & 
\texttt{class Shl a b c where (<{}<) : a -> b -> c}
 \\

\texttt{a >{}> b}
 & 
\texttt{a.shr(b)}
 & 
std::ops::Shr<b,c>
 & 
\texttt{class Shr a b c where (>{}>) : a -> b -> c}
 \\
\end{longtable*}

The \texttt{for} loop uses the special trait \texttt{std::iter::Iterator}, as follows:
\begin{quote}{\ttfamily \raggedright \noindent
\DUrole{c1}{//~Rust\\
}~\\
\DUrole{c1}{//~the~following~expression:\\
}\DUrole{p}{{[}}\DUrole{o}{<}\DUrole{n}{label}\DUrole{o}{>}\DUrole{p}{{]}}~\DUrole{k}{for}~\DUrole{o}{<}\DUrole{n}{pat}\DUrole{o}{>}~\DUrole{n}{in}~\DUrole{o}{<}\DUrole{n}{iterator}~\DUrole{n}{expression}\DUrole{o}{>}~\DUrole{p}{\{}~\\
~~~\DUrole{n}{body}\DUrole{p}{...;}~\\
\DUrole{p}{\}}~\\
~\\
\DUrole{c1}{//~...~expands~to~(internally):\\
}\DUrole{k}{match}~\DUrole{o}{\&}\DUrole{k}{mut}~\DUrole{o}{<}\DUrole{n}{iterator}~\DUrole{n}{expression}\DUrole{o}{>}~\DUrole{p}{\{}~\\
~~\DUrole{n}{\_v}~\DUrole{o}{=>}~\DUrole{p}{{[}}\DUrole{n}{label}\DUrole{p}{{]}}~\DUrole{k}{loop}~\DUrole{p}{\{}~\\
~~~~\DUrole{k}{match}~\DUrole{n}{\_v}\DUrole{p}{.}\DUrole{n}{next}\DUrole{p}{()}~\DUrole{p}{\{}~\\
~~~~~~~\DUrole{n}{None}~\DUrole{o}{=>}~\DUrole{k}{break}\DUrole{p}{,}~\\
~~~~~~~\DUrole{n}{Some}\DUrole{p}{(}\DUrole{o}{<}\DUrole{n}{pat}\DUrole{o}{>}\DUrole{p}{)}~\DUrole{o}{=>}~\DUrole{p}{\{}~\DUrole{n}{body}\DUrole{p}{...}~\DUrole{p}{\}}~\\
~~~~\DUrole{p}{\}}~\\
~~\DUrole{p}{\}}~\\
\DUrole{p}{\}}
}
\end{quote}

The method \texttt{next} is implemented by \texttt{Iterator}.  The return type
of \texttt{next} is \texttt{Option}, which can have values \texttt{None}, to mean
“nothing left to iterate”, or \texttt{Some(x)}, to mean the next iteration
value is \texttt{x}.

\section{Ad-hoc objects and methods%
  \label{ad-hoc-objects-and-methods}%
}

In addition to the mechanism offered by traits, any \texttt{struct} or
\texttt{enum} can be \emph{decorated} with one or more method interface(s)
using “\texttt{impl}”, separately from its definition and/or in different
modules:

\setlength{\DUtablewidth}{\linewidth}
\begin{longtable*}[c]{p{0.470\DUtablewidth}p{0.470\DUtablewidth}}

{\ttfamily \raggedright \noindent
\DUrole{c1}{//~Rust\\
}\DUrole{k}{struct}~\DUrole{n}{R}~\DUrole{p}{\{}\DUrole{n}{x}\DUrole{o}{:}\DUrole{k}{int}\DUrole{p}{\};}~\\
~\\
\DUrole{c1}{//~in~some~module:\\
}\DUrole{k}{impl}~\DUrole{n}{R}~\DUrole{p}{\{}~\\
~\DUrole{k}{fn}~\DUrole{n}{hello}\DUrole{p}{(}\DUrole{o}{\&}\DUrole{n}{self}\DUrole{p}{)}~\DUrole{p}{\{}~\\
~~\DUrole{n}{println}\DUrole{o}{!}\DUrole{p}{(}\DUrole{s}{\textquotedbl{}hello~\{\}\textquotedbl{}}\DUrole{p}{,}~\DUrole{n}{self}\DUrole{p}{.}\DUrole{n}{x}\DUrole{p}{);}~\\
~\\
~\\
~\\
~\DUrole{p}{\}}~\\
\DUrole{p}{\}}~\\
\DUrole{c1}{//~possibly~somewhere~else:\\
}\DUrole{k}{impl}~\DUrole{n}{R}~\DUrole{p}{\{}~\\
~\DUrole{k}{fn}~\DUrole{n}{world}\DUrole{p}{(}\DUrole{o}{\&}\DUrole{n}{self}\DUrole{p}{)}~\DUrole{p}{\{}~\\
~~\DUrole{n}{println}\DUrole{o}{!}\DUrole{p}{(}\DUrole{s}{\textquotedbl{}world\textquotedbl{}}\DUrole{p}{);}~\\
~\DUrole{p}{\}}~\\
\DUrole{p}{\}}~\\
\DUrole{c1}{//~Example~use:\\
}\DUrole{k}{fn}~\DUrole{n}{main}\DUrole{p}{()}~\DUrole{p}{\{}~\\
~~\DUrole{k}{let}~\DUrole{n}{v}~\DUrole{o}{=}~\DUrole{n}{R}~\DUrole{p}{\{}\DUrole{n}{x}\DUrole{o}{:}\DUrole{m}{10}\DUrole{k}{}\DUrole{p}{\};}~\\
~~\DUrole{n}{v}\DUrole{p}{.}\DUrole{n}{hello}\DUrole{p}{();}~\\
~~\DUrole{n}{v}\DUrole{p}{.}\DUrole{n}{world}\DUrole{p}{();}~\\
~\\
~~\DUrole{p}{(}\DUrole{n}{R}\DUrole{p}{\{}\DUrole{n}{x}\DUrole{o}{:}\DUrole{m}{20}\DUrole{k}{}\DUrole{p}{\}).}\DUrole{n}{hello}\DUrole{p}{();}~\\
\DUrole{p}{\}}
}
 & 
{\ttfamily \raggedright \noindent
\DUrole{c1}{//~Rust\\
}\DUrole{k}{enum}~\DUrole{n}{E}~\DUrole{p}{\{}\DUrole{n}{A}\DUrole{p}{,}~\DUrole{n}{B}\DUrole{p}{\};}~\\
~\\
\DUrole{c1}{//~in~some~module:\\
}\DUrole{k}{impl}~\DUrole{n}{E}~\DUrole{p}{\{}~\\
~\DUrole{k}{fn}~\DUrole{n}{hello}\DUrole{p}{(}\DUrole{o}{\&}\DUrole{n}{self}\DUrole{p}{)}~\DUrole{p}{\{}~\\
~~~\DUrole{k}{match}~\DUrole{n}{self}~\DUrole{p}{\{}~\\
~~~~\DUrole{o}{\&}\DUrole{n}{A}~\DUrole{o}{=>}~\DUrole{n}{println}\DUrole{o}{!}\DUrole{p}{(}\DUrole{s}{\textquotedbl{}hello~A\textquotedbl{}}\DUrole{p}{),}~\\
~~~~\DUrole{o}{\&}\DUrole{n}{B}~\DUrole{o}{=>}~\DUrole{n}{println}\DUrole{o}{!}\DUrole{p}{(}\DUrole{s}{\textquotedbl{}hello~B\textquotedbl{}}\DUrole{p}{)}~\\
~~~\DUrole{p}{\}}~\\
~\DUrole{p}{\}}~\\
\DUrole{p}{\}}~\\
\DUrole{c1}{//~possibly~somewhere~else:\\
}\DUrole{k}{impl}~\DUrole{n}{E}~\DUrole{p}{\{}~\\
~\DUrole{k}{fn}~\DUrole{n}{world}\DUrole{p}{(}\DUrole{o}{\&}\DUrole{n}{self}\DUrole{p}{)}~\DUrole{p}{\{}~\\
~~\DUrole{n}{println}\DUrole{o}{!}\DUrole{p}{(}\DUrole{s}{\textquotedbl{}world\textquotedbl{}}\DUrole{p}{);}~\\
~\DUrole{p}{\}}~\\
\DUrole{p}{\}}~\\
\DUrole{c1}{//~Example~use:\\
}\DUrole{k}{fn}~\DUrole{n}{main}\DUrole{p}{()}~\DUrole{p}{\{}~\\
~~\DUrole{k}{let}~\DUrole{n}{v}~\DUrole{o}{=}~\DUrole{n}{A}\DUrole{p}{;}~\\
~~\DUrole{n}{v}\DUrole{p}{.}\DUrole{n}{hello}\DUrole{p}{();}~\\
~~\DUrole{n}{v}\DUrole{p}{.}\DUrole{n}{world}\DUrole{p}{();}~\\
~\\
~~\DUrole{n}{B}\DUrole{p}{.}\DUrole{n}{hello}\DUrole{p}{();}~\\
\DUrole{p}{\}}
}
 \\
\end{longtable*}

\section{Safe references%
  \label{safe-references}%
}

Like in C, by default function parameters in Rust are passed
by value. For large data types, copying the data may be
expensive so one may want to use \emph{references} instead.

For any object \texttt{v} of type \texttt{T}, it is possible to create a
\emph{reference} to that object using the expression “\texttt{\&v}”.
The reference itself then has type \texttt{\&T}.
\begin{quote}{\ttfamily \raggedright \noindent
\DUrole{k}{use}~\DUrole{n}{std}\DUrole{o}{::}\DUrole{n}{num}\DUrole{o}{::}\DUrole{n}{sqrt}\DUrole{p}{;}~\\
\DUrole{k}{struct}~\DUrole{n}{Pt}~\DUrole{p}{\{}~\DUrole{n}{x}\DUrole{o}{:}\DUrole{k}{f32}\DUrole{p}{,}~\DUrole{n}{y}\DUrole{o}{:}\DUrole{k}{f32}~\DUrole{p}{\}}~\\
~\\
\DUrole{c1}{//~This~works,~but~may~be~expensive:\\
}\DUrole{k}{fn}~\DUrole{n}{dist1}\DUrole{p}{(}\DUrole{n}{p1}~\DUrole{o}{:}~\DUrole{n}{Pt}\DUrole{p}{,}~\DUrole{n}{p2}\DUrole{o}{:}~\DUrole{n}{Pt}\DUrole{p}{)}~\DUrole{o}{->}~\DUrole{k}{f32}~\DUrole{p}{\{}~\\
~~~\DUrole{k}{let}~\DUrole{n}{xd}~\DUrole{o}{=}~\DUrole{n}{p1}\DUrole{p}{.}\DUrole{n}{x}~\DUrole{o}{-}~\DUrole{n}{p2}\DUrole{p}{.}\DUrole{n}{x}\DUrole{p}{;}~\\
~~~\DUrole{k}{let}~\DUrole{n}{yd}~\DUrole{o}{=}~\DUrole{n}{p1}\DUrole{p}{.}\DUrole{n}{y}~\DUrole{o}{-}~\DUrole{n}{p2}\DUrole{p}{.}\DUrole{n}{y}\DUrole{p}{;}~\\
~~~\DUrole{n}{sqrt}\DUrole{p}{(}\DUrole{n}{xd}~\DUrole{o}{*}~\DUrole{n}{xd}~\DUrole{o}{+}~\DUrole{n}{yd}~\DUrole{o}{*}~\DUrole{n}{yd}\DUrole{p}{)}~\\
\DUrole{p}{\}}~\\
~\\
\DUrole{c1}{//~Same,~using~references:\\
}\DUrole{k}{fn}~\DUrole{n}{dist2}\DUrole{p}{(}\DUrole{n}{p1}~\DUrole{o}{:}~\DUrole{o}{\&}\DUrole{n}{Pt}\DUrole{p}{,}~\DUrole{n}{p2}\DUrole{o}{:}~\DUrole{o}{\&}\DUrole{n}{Pt}\DUrole{p}{)}~\DUrole{o}{->}~\DUrole{k}{f32}~\DUrole{p}{\{}~\\
~~~\DUrole{k}{let}~\DUrole{n}{xd}~\DUrole{o}{=}~\DUrole{n}{p1}\DUrole{p}{.}\DUrole{n}{x}~\DUrole{o}{-}~\DUrole{n}{p2}\DUrole{p}{.}\DUrole{n}{x}\DUrole{p}{;}~\\
~~~\DUrole{k}{let}~\DUrole{n}{yd}~\DUrole{o}{=}~\DUrole{n}{p1}\DUrole{p}{.}\DUrole{n}{y}~\DUrole{o}{-}~\DUrole{n}{p2}\DUrole{p}{.}\DUrole{n}{y}\DUrole{p}{;}~\\
~~~\DUrole{n}{sqrt}\DUrole{p}{(}\DUrole{n}{xd}~\DUrole{o}{*}~\DUrole{n}{xd}~\DUrole{o}{+}~\DUrole{n}{yd}~\DUrole{o}{*}~\DUrole{n}{yd}\DUrole{p}{)}~\\
\DUrole{p}{\}}~\\
~\\
\DUrole{c1}{//~Usage:\\
}\DUrole{k}{fn}~\DUrole{n}{main}\DUrole{p}{()}~\DUrole{p}{\{}~\\
~~~\DUrole{k}{let}~\DUrole{n}{a}~\DUrole{o}{=}~\DUrole{n}{Pt}~\DUrole{p}{\{}~\DUrole{n}{x}\DUrole{o}{:}\DUrole{m}{1.0}\DUrole{k}{}\DUrole{p}{,}~\DUrole{n}{y}\DUrole{o}{:}\DUrole{m}{2.0}\DUrole{k}{}~\DUrole{p}{\};}~\\
~~~\DUrole{k}{let}~\DUrole{n}{b}~\DUrole{o}{=}~\DUrole{n}{Pt}~\DUrole{p}{\{}~\DUrole{n}{x}\DUrole{o}{:}\DUrole{m}{0.0}\DUrole{k}{}\DUrole{p}{,}~\DUrole{n}{y}\DUrole{o}{:}\DUrole{m}{3.0}\DUrole{k}{}~\DUrole{p}{\};}~\\
~~~\DUrole{n}{println}\DUrole{o}{!}\DUrole{p}{(}\DUrole{s}{\textquotedbl{}\{\}\textquotedbl{}}\DUrole{p}{,}~\DUrole{n}{dist1}\DUrole{p}{(}\DUrole{n}{a}\DUrole{p}{,}~\DUrole{n}{b}\DUrole{p}{));}~\\
~~~\DUrole{n}{println}\DUrole{o}{!}\DUrole{p}{(}\DUrole{s}{\textquotedbl{}\{\}\textquotedbl{}}\DUrole{p}{,}~\DUrole{n}{dist2}\DUrole{p}{(}\DUrole{o}{\&}\DUrole{n}{a}\DUrole{p}{,}~\DUrole{o}{\&}\DUrole{n}{b}\DUrole{p}{));}~\\
\DUrole{p}{\}}
}
\end{quote}

As illustrated in this example, Rust provides some syntactic
sugar for \texttt{struct} references: it is possible to write \texttt{p1.x} if
\texttt{p1} is of type \texttt{\&Pt} and \texttt{Pt} has a field named \texttt{x}. This
sugar is also available for method calls (\texttt{x.foo()}).

However, in many other cases, either the referenced value must be “retrieved”
using the unary \texttt{*} operator, or patterns must be matched using \texttt{\&}:
\begin{quote}{\ttfamily \raggedright \noindent
\DUrole{k}{fn}~\DUrole{n}{add3}\DUrole{p}{(}\DUrole{n}{x}~\DUrole{o}{:}~\DUrole{o}{\&}\DUrole{k}{int}\DUrole{p}{)}~\DUrole{o}{->}~\DUrole{k}{int}~\DUrole{p}{\{}~\\
~~~\DUrole{n}{println}\DUrole{o}{!}\DUrole{p}{(}\DUrole{s}{\textquotedbl{}x~:~\{\}\textquotedbl{}}\DUrole{p}{,}~\DUrole{o}{*}\DUrole{n}{x}\DUrole{p}{);}~\DUrole{c1}{//~OK\\
}~~~\DUrole{m}{3}\DUrole{k}{}~\DUrole{o}{+}~\DUrole{n}{x}\DUrole{p}{;}~\DUrole{c1}{//~invalid!~(+~cannot~apply~to~\&int)\\
}~~~\DUrole{m}{3}\DUrole{k}{}~\DUrole{o}{+}~\DUrole{o}{*}\DUrole{n}{x}\DUrole{p}{;}~\DUrole{c1}{//~OK\\
}\DUrole{p}{\}}~\\
~\\
\DUrole{k}{fn}~\DUrole{n}{test}\DUrole{o}{<}\DUrole{n}{t}\DUrole{o}{>}\DUrole{p}{(}\DUrole{n}{x}~\DUrole{o}{:}~\DUrole{o}{\&}\DUrole{n}{Option}\DUrole{o}{<}\DUrole{n}{t}\DUrole{o}{>}\DUrole{p}{)}~\DUrole{o}{->}~\DUrole{n}{bool}~\DUrole{p}{\{}~\\
~~~\DUrole{k}{match}~\DUrole{o}{*}\DUrole{n}{x}~\DUrole{p}{\{}~\\
~~~~~\DUrole{n}{None}~~~~~\DUrole{o}{=>}~\DUrole{k}{false}\DUrole{p}{,}~\\
~~~~~\DUrole{n}{Some}\DUrole{p}{(}\DUrole{n}{\_}\DUrole{p}{)}~~\DUrole{o}{=>}~\DUrole{k}{true}~\\
~~~\DUrole{p}{\}}~\\
~~~\DUrole{c1}{//~Also~valid,~equivalent:\\
}~~~\DUrole{k}{match}~\DUrole{n}{x}~\DUrole{p}{\{}~\\
~~~~~\DUrole{o}{\&}\DUrole{n}{None}~~~~\DUrole{o}{=>}~\DUrole{k}{false}\DUrole{p}{,}~\\
~~~~~\DUrole{o}{\&}\DUrole{n}{Some}\DUrole{p}{(}\DUrole{n}{\_}\DUrole{p}{)}~\DUrole{o}{=>}~\DUrole{k}{true}~\\
~~~\DUrole{p}{\}}~\\
\DUrole{p}{\}}
}
\end{quote}

Simple references not not allow modifying the underlying object;
if mutability by reference is desired, use “\texttt{\&mut}” as follows:
\begin{quote}{\ttfamily \raggedright \noindent
\DUrole{k}{fn}~\DUrole{n}{incx}\DUrole{p}{(}\DUrole{n}{x}~\DUrole{o}{:}~\DUrole{o}{\&}\DUrole{k}{int}\DUrole{p}{)}~\DUrole{p}{\{}~\\
~~~\DUrole{n}{x}~\DUrole{o}{=}~\DUrole{n}{x}~\DUrole{o}{+}~\DUrole{m}{1}\DUrole{k}{}\DUrole{p}{;}~~~\DUrole{c1}{//~invalid!~(mismatched~types~in~\textquotedbl{}int\&~=~int\textquotedbl{})\\
}~~~\DUrole{o}{*}\DUrole{n}{x}~\DUrole{o}{=}~\DUrole{o}{*}\DUrole{n}{x}~\DUrole{o}{+}~\DUrole{m}{1}\DUrole{k}{}\DUrole{p}{;}~\DUrole{c1}{//~invalid!~(*x~is~immutable)\\
}\DUrole{p}{\}}~\\
~\\
\DUrole{k}{fn}~\DUrole{n}{inc}\DUrole{p}{(}\DUrole{n}{x}~\DUrole{o}{:}~\DUrole{o}{\&}\DUrole{k}{mut}~\DUrole{k}{int}\DUrole{p}{)}~\DUrole{p}{\{}~\\
~~~\DUrole{n}{x}~\DUrole{o}{=}~\DUrole{n}{x}~\DUrole{o}{+}~\DUrole{m}{1}\DUrole{k}{}\DUrole{p}{;}~~~\DUrole{c1}{//~invalid!~(x~is~immutable)\\
}~~~\DUrole{o}{*}\DUrole{n}{x}~\DUrole{o}{=}~\DUrole{o}{*}\DUrole{n}{x}~\DUrole{o}{+}~\DUrole{m}{1}\DUrole{k}{}\DUrole{p}{;}~\DUrole{c1}{//~OK\\
}\DUrole{p}{\}}~\\
~\\
\DUrole{k}{fn}~\DUrole{n}{main}\DUrole{p}{()}~\DUrole{p}{\{}~\\
~~~\DUrole{n}{inc}\DUrole{p}{(}\DUrole{o}{\&}\DUrole{m}{3}\DUrole{k}{}\DUrole{p}{);}~\DUrole{c1}{//~invalid!~(3~is~immutable)\\
}~~~\DUrole{n}{inc}\DUrole{p}{(}\DUrole{o}{\&}\DUrole{k}{mut}~\DUrole{m}{3}\DUrole{k}{}\DUrole{p}{);}~\DUrole{c1}{//~OK,~temp~var~forgotten~after~call\\
}~~~\DUrole{k}{let}~\DUrole{n}{v}~\DUrole{o}{=}~\DUrole{m}{3}\DUrole{k}{}\DUrole{p}{;}~\\
~~~\DUrole{n}{inc}\DUrole{p}{(}\DUrole{o}{\&}\DUrole{n}{v}\DUrole{p}{);}~\DUrole{c1}{//~invalid!~(v~is~immutable)\\
}~~~\DUrole{n}{inc}\DUrole{p}{(}\DUrole{o}{\&}\DUrole{k}{mut}~\DUrole{n}{v}\DUrole{p}{);}~\DUrole{c1}{//~OK,~temp~var~forgotten~after~call\\
}~~~\DUrole{k}{let}~\DUrole{k}{mut}~\DUrole{n}{w}~\DUrole{o}{=}~\DUrole{m}{3}\DUrole{k}{}\DUrole{p}{;}~\\
~~~\DUrole{n}{inc}\DUrole{p}{(}\DUrole{o}{\&}\DUrole{n}{w}\DUrole{p}{);}~\DUrole{c1}{//~OK\\
}\DUrole{p}{\}}
}
\end{quote}

Rust's type system \emph{forbids mutable aliases} via references: it is not
possible to modify the same object using different names via
references, unlike in C. This is done via the concept of \textbf{borrowing}: while
ownership of an object is borrowed by a reference, the original variable
cannot be used any more. For example:
\begin{quote}{\ttfamily \raggedright \noindent
~\DUrole{k}{fn}~\DUrole{n}{main}\DUrole{p}{()}~\DUrole{p}{\{}~\\
~~~~\DUrole{k}{let}~\DUrole{k}{mut}~\DUrole{n}{a}~\DUrole{o}{=}~\DUrole{m}{1}\DUrole{k}{}\DUrole{p}{;}~\\
~~~~\DUrole{n}{a}~\DUrole{o}{=}~\DUrole{m}{2}\DUrole{k}{}\DUrole{p}{;}~\DUrole{c1}{//~OK\\
}~~~~\DUrole{n}{println}\DUrole{o}{!}\DUrole{p}{(}\DUrole{s}{\textquotedbl{}\{\}\textquotedbl{}}\DUrole{p}{,}~\DUrole{n}{a}\DUrole{p}{);}~\DUrole{c1}{//~OK,~prints~2\\
}~~~~\DUrole{c1}{//~\{...\}~introduces~a~new~scope:\\
}~~~~\DUrole{p}{\{}~\\
~~~~~~\DUrole{k}{let}~\DUrole{n}{ra}~\DUrole{o}{=}~\DUrole{o}{\&}\DUrole{k}{mut}~\DUrole{n}{a}\DUrole{p}{;}~\\
~~~~~~\DUrole{n}{a}~\DUrole{o}{=}~\DUrole{m}{3}\DUrole{k}{}\DUrole{p}{;}~~~\DUrole{c1}{//~invalid!~(a~is~borrowed~by~ra)\\
}~~~~~~\DUrole{o}{*}\DUrole{n}{ra}~\DUrole{o}{=}~\DUrole{m}{3}\DUrole{k}{}\DUrole{p}{;}~\DUrole{c1}{//~OK\\
}~~~~~~\DUrole{n}{println}\DUrole{o}{!}\DUrole{p}{(}\DUrole{s}{\textquotedbl{}\{\}\textquotedbl{}}\DUrole{p}{,}~\DUrole{n}{a}\DUrole{p}{);}~\DUrole{c1}{//~invalid!~(a~is~borrowed)\\
}~~~~~~\DUrole{n}{println}\DUrole{o}{!}\DUrole{p}{(}\DUrole{s}{\textquotedbl{}\{\}\textquotedbl{}}\DUrole{p}{,}~\DUrole{o}{*}\DUrole{n}{ra}\DUrole{p}{);}~\DUrole{c1}{//~OK,~prints~3\\
}~~~~\DUrole{p}{\};}~\\
~~~~\DUrole{c1}{//~end~of~scope,~rb~is~dropped.\\
}~~~~\DUrole{n}{println}\DUrole{o}{!}\DUrole{p}{(}\DUrole{s}{\textquotedbl{}\{\}\textquotedbl{}}\DUrole{p}{,}~\DUrole{n}{a}\DUrole{p}{);}~\DUrole{c1}{//~OK,~a~not~borrowed~any~more~(prints~3)\\
}\DUrole{p}{\}}
}
\end{quote}

Reference types, together with “all values are immutable by default”
and the ownership/borrowing rules, are the core features of Rust's
type system that make it fundamentally safer than C's.

\section{Lifetime and storage, and managed objects%
  \label{lifetime-and-storage-and-managed-objects}%
}

\DUadmonition[note]{
\DUtitle[note]{Note}

The concepts and syntax presented in this section are new in Rust
version 0.11. Rust 0.10 and previous versions used different
concepts, terminology and syntax.

Also, at the time of this writing (July 2014) the official Rust
manual and tutorials are not yet updated to the latest language
implementation. This section follows the implementation, not the
manuals.
}

\subsection{Introduction%
  \label{introduction}%
}

Rust is defined over the same \emph{abstract machine model} as C.  The
abstract machine has segments of memory, and the language's run-time
machinery can allocate and deallocate segments of memory over time during
program execution.

In the abstract machine, the two following concepts are defined in
Rust just like in C:
\begin{itemize}

\item the \emph{storage} of an object determines the type of memory where the object is stored;

\item the \emph{lifetime} or \emph{storage duration} of an object is the segment of
time from the point an object is allocated to the point where it is
deallocated.

\end{itemize}

All memory-related problems in C come from the fact that C programs
can manipulate references to objects \emph{outside of their lifetime} (ie. before
they are allocated or after they are deallocated), or \emph{outside of their
storage} (ie. at lower or higher addresses in memory).

Rust goes to great lengths to prevent all these problems altogether,
by \textbf{ensuring that objects cannot be used outside of their lifetime or
their storage}.

\subsection{Storages in Rust vs. C%
  \label{storages-in-rust-vs-c}%
}

There are four kinds of storage in the C/Rust abstract machine:
static, thread, automatic and allocated.

\setlength{\DUtablewidth}{\linewidth}
\begin{longtable*}[c]{p{0.133\DUtablewidth}p{0.354\DUtablewidth}p{0.400\DUtablewidth}}
\textbf{%
Storage
} & \textbf{%
Type of memory used
} & \textbf{%
Example in C
} \\
\endfirsthead
\textbf{%
Storage
} & \textbf{%
Type of memory used
} & \textbf{%
Example in C
} \\
\endhead
\multicolumn{3}{c}{\hfill ... continued on next page} \\
\endfoot
\endlastfoot

static
 & 
process-wide special segment
 & 
\texttt{static int i;}
 \\

automatic
 & 
stack
 & 
\texttt{int i;} (in function scope)
 \\

allocated
 & 
global heap
 & 
\texttt{int *i = malloc(sizeof(int));}
 \\

thread
 & 
per-thread heap
 & 
\texttt{\_Thread\_local int i;}
 \\
\end{longtable*}

In C, the lifetime of an object is solely determined by its storage:

\setlength{\DUtablewidth}{\linewidth}
\begin{longtable*}[c]{p{0.133\DUtablewidth}p{0.726\DUtablewidth}}
\textbf{%
Storage
} & \textbf{%
Lifetime in C
} \\
\endfirsthead
\textbf{%
Storage
} & \textbf{%
Lifetime in C
} \\
\endhead
\multicolumn{2}{c}{\hfill ... continued on next page} \\
\endfoot
\endlastfoot

static
 & 
From program start to termination
 \\

automatic
 & 
From start of scope to end of scope, one per activation frame
 \\

allocated
 & 
From \texttt{malloc} to \texttt{free} (or \texttt{mmap} to \texttt{munmap}, etc)
 \\

thread
 & 
From object allocation to thread termination
 \\
\end{longtable*}

In Rust, the lifetime of static and automatic objects
is the same as in C; however:
\begin{itemize}

\item Rust introduces a new “box” type with dedicated syntax for heap
allocated objects, which are called \emph{managed objects}.

Rust supports multiple management strategies for boxes, \emph{associated
to different typing rules}.

\item The default management strategy for boxes ensures that \textbf{boxes are
uniquely owned}, so that the compiler knows precisely when the
lifetime of a box ends and where it can be safely deallocated
without the need for extra machinery like reference counting or
garbage collection.

\item Another strategy is GC, that uses deferred garbage collection: the
storage for GC boxes is reclaimed and made available for reuse at
some point when the Rust run-time system can determine that they are
not needed any more. (This may delay reclamation for an unpredictable
amount of time.) References to GC boxes need not be unique, so GC boxes
are an appropriate type to build complex data structures with common
nodes, like arbitrary graphs.

\item Unlike C, \textbf{Rust forbids sharing of managed objects across
threads}: like in Erlang, objects have a single owner for
allocation, and most objects are entirely private to each
thread. This eliminates most data races. Also the single thread ownership
implies that garbage collection for GC objects does not require
inter-thread synchronization, so it is easier to implement and can
run faster.

\item As with references to normal objects, \textbf{Rust forbids mutable
aliases} with references to managed objects.

\item \textbf{All vector accesses are bound checked, and references do not support
pointer arithmetic.}

\item Rust strongly discourages all uses of unmanaged pointers, by
tainting them with the \texttt{unsafe} attribute which systematically
elicits compilers warnings.

\end{itemize}

\subsection{Creating and using boxes%
  \label{creating-and-using-boxes}%
}

The Rust handling of managed objects is relatively simple: the \texttt{box}
keyword in expressions “puts objects into boxes”, where their lifetime
is managed dynamically. The immutability and ownership of boxes is
handled like with references described earlier:
\begin{quote}{\ttfamily \raggedright \noindent
\DUrole{k}{fn}~\DUrole{n}{main}\DUrole{p}{()}~\DUrole{p}{\{}~\\
~\\
~~\DUrole{k}{let}~\DUrole{n}{b}~\DUrole{o}{=}~\DUrole{n}{box}~\DUrole{m}{3}\DUrole{k}{i}\DUrole{p}{;}~\DUrole{c1}{//~b~has~type~Box<int>\\
}~~\DUrole{n}{b}~\DUrole{o}{=}~\DUrole{m}{4}\DUrole{k}{i}\DUrole{p}{;}~\DUrole{c1}{//~invalid!~can't~assign~int~to~box<int>\\
}~~\DUrole{o}{*}\DUrole{n}{b}~\DUrole{o}{=}~\DUrole{m}{4}\DUrole{k}{i}\DUrole{p}{;}~\DUrole{c1}{//~invalid!~b~is~immutable,~so~is~the~box\\
}~~\DUrole{k}{let}~\DUrole{k}{mut}~\DUrole{n}{c}~\DUrole{o}{=}~\DUrole{n}{box}~\DUrole{m}{3}\DUrole{k}{i}\DUrole{p}{;}~\DUrole{c1}{//~c~has~type~mut~Box<int>\\
}~~\DUrole{o}{*}\DUrole{n}{c}~\DUrole{o}{=}~\DUrole{m}{4}\DUrole{k}{i}\DUrole{p}{;}~\DUrole{c1}{//~OK\\
}~\\
~~\DUrole{k}{let}~\DUrole{n}{v}~\DUrole{o}{=}~\DUrole{n}{box}~\DUrole{n}{vec}\DUrole{o}{!}\DUrole{p}{(}\DUrole{m}{1}\DUrole{k}{i}\DUrole{p}{,}\DUrole{m}{2}\DUrole{k}{}\DUrole{p}{,}\DUrole{m}{3}\DUrole{k}{}\DUrole{p}{);}~\DUrole{c1}{//~r1~has~type~Box<Vec<int>{}>\\
}~~\DUrole{o}{*}\DUrole{n}{v}\DUrole{p}{.}\DUrole{n}{get\_mut}\DUrole{p}{(}\DUrole{m}{0}\DUrole{k}{}\DUrole{p}{)}~\DUrole{o}{=}~\DUrole{m}{42}\DUrole{k}{}\DUrole{p}{;}~\DUrole{c1}{//~invalid!~v~is~immutable,~so~is~the~box\\
}~~\DUrole{k}{let}~\DUrole{k}{mut}~\DUrole{n}{w}~\DUrole{o}{=}~\DUrole{n}{box}~\DUrole{n}{vec}\DUrole{o}{!}\DUrole{p}{(}\DUrole{m}{1}\DUrole{k}{i}\DUrole{p}{,}\DUrole{m}{2}\DUrole{k}{}\DUrole{p}{,}\DUrole{m}{3}\DUrole{k}{}\DUrole{p}{);}~\\
~~\DUrole{o}{*}\DUrole{n}{w}\DUrole{p}{.}\DUrole{n}{get\_mut}\DUrole{p}{(}\DUrole{m}{0}\DUrole{k}{}\DUrole{p}{)}~\DUrole{o}{=}~\DUrole{m}{42}\DUrole{k}{}\DUrole{p}{;}~\DUrole{c1}{//~OK,~rust~0.11.1+~also~permits~w{[}0{]}~=~42\\
}~\\
~~\DUrole{k}{let}~\DUrole{n}{z}~\DUrole{o}{=}~\DUrole{n}{w}\DUrole{p}{;}~\DUrole{c1}{//~z~has~type~Box<Vec<int>{}>,~captures~box\\
}~~\DUrole{n}{println}\DUrole{o}{!}\DUrole{p}{(}\DUrole{s}{\textquotedbl{}\{\}\textquotedbl{}}\DUrole{p}{,}~\DUrole{n}{w}\DUrole{p}{);}~\DUrole{c1}{//~invalid!~box~has~moved~to~z\\
}~\\
~~\DUrole{n}{w}~\DUrole{o}{=}~\DUrole{n}{box}~\DUrole{n}{vec}\DUrole{o}{!}\DUrole{p}{(}\DUrole{m}{2}\DUrole{k}{i}\DUrole{p}{,}\DUrole{m}{4}\DUrole{k}{}\DUrole{p}{,}\DUrole{m}{5}\DUrole{k}{}\DUrole{p}{);}~\DUrole{c1}{//~overwrites~reference,~not~box~contents\\
}~~\DUrole{n}{println}\DUrole{o}{!}\DUrole{p}{(}\DUrole{s}{\textquotedbl{}\{\}~\{\}\textquotedbl{}}\DUrole{p}{,}~\DUrole{n}{w}\DUrole{p}{,}~\DUrole{n}{z}\DUrole{p}{);}~\DUrole{c1}{//~OK,~prints~{[}2,4,5{]}~{[}42,2,3{]}\\
}\DUrole{p}{\}}
}
\end{quote}

The \texttt{box} keyword in expressions is actually a shorthand form for \texttt{box(HEAP)}.
The general form “\texttt{box(A) E}” places the result of the evaluation of \texttt{E}
in a memory object allocated by \texttt{A}, which is a trait. As of Rust 0.11, the only other
allocator in the standard library is \texttt{GC}, for garbage collected objects.

\setlength{\DUtablewidth}{\linewidth}
\begin{longtable*}[c]{p{0.145\DUtablewidth}p{0.125\DUtablewidth}p{0.675\DUtablewidth}}
\textbf{%
Expression
} & \textbf{%
Type
} & \textbf{%
Meaning
} \\
\endfirsthead
\textbf{%
Expression
} & \textbf{%
Type
} & \textbf{%
Meaning
} \\
\endhead
\multicolumn{3}{c}{\hfill ... continued on next page} \\
\endfoot
\endlastfoot

\texttt{box v}
 & 
Box<T>
 & 
Unique reference to a copy of \texttt{v}, shorthand for \texttt{box(HEAP) v}
 \\

\texttt{box(GC) v}
 & 
Gc<T>
 & 
Garbage-collected smart pointer to a copy of \texttt{v}
 \\

\texttt{box(A) v}
 & 
??<T>
 & 
Some kind of smart pointer; \texttt{A} must implement a special trait.
 \\
\end{longtable*}

\subsection{Recursive data structures%
  \label{recursive-data-structures}%
}

Managed objects are the “missing link” to implement proper recursive
algebraic data types:

\setlength{\DUtablewidth}{\linewidth}
\begin{longtable*}[c]{p{0.315\DUtablewidth}p{0.315\DUtablewidth}p{0.315\DUtablewidth}}

{\ttfamily \raggedright \noindent
\DUrole{c1}{//~Rust\\
}\DUrole{k}{enum}~\DUrole{n}{Lst}\DUrole{o}{<}\DUrole{n}{t}\DUrole{o}{>}~\DUrole{p}{\{}~\\
~~~\DUrole{n}{Nil}\DUrole{p}{,}~\\
~~~\DUrole{n}{Cons}\DUrole{p}{(}\DUrole{n}{t}\DUrole{p}{,}~\DUrole{n}{Box}\DUrole{o}{<}\DUrole{n}{Lst}\DUrole{o}{<}\DUrole{n}{t}\DUrole{o}{>{}>}\DUrole{p}{)}~\\
\DUrole{p}{\}}~\\
\DUrole{k}{fn}~\DUrole{n}{len}\DUrole{o}{<}\DUrole{n}{t}\DUrole{o}{>}\DUrole{p}{(}\DUrole{n}{l}~\DUrole{o}{:}~\DUrole{o}{\&}\DUrole{n}{Lst}\DUrole{o}{<}\DUrole{n}{t}\DUrole{o}{>}\DUrole{p}{)}~\DUrole{o}{->}~\DUrole{k}{uint}~\DUrole{p}{\{}~\\
~~~\DUrole{k}{match}~\DUrole{o}{*}\DUrole{n}{l}~\DUrole{p}{\{}~\\
~~~~~\DUrole{n}{Nil}~\DUrole{o}{=>}~\DUrole{m}{0}\DUrole{k}{}\DUrole{p}{,}~\\
~~~~~\DUrole{n}{Cons}\DUrole{p}{(}\DUrole{n}{\_}\DUrole{p}{,}~\DUrole{k}{ref}~\DUrole{n}{x}\DUrole{p}{)}~\DUrole{o}{=>}~\DUrole{m}{1}\DUrole{k}{}~\DUrole{o}{+}~\DUrole{n}{len}\DUrole{p}{(}\DUrole{o}{*}\DUrole{n}{x}\DUrole{p}{)}~\\
~~~\DUrole{p}{\}}~\\
\DUrole{p}{\}}~\\
~\\
\DUrole{k}{fn}~\DUrole{n}{main}\DUrole{p}{()}~\DUrole{p}{\{}~\\
~~~\DUrole{k}{let}~\DUrole{n}{l}~\DUrole{o}{=}~\DUrole{n}{box}~\DUrole{n}{Cons}\DUrole{p}{(}\DUrole{sc}{'a'}\DUrole{p}{,}~\\
~~~~~~~~~~~~\DUrole{n}{box}~\DUrole{n}{Cons}\DUrole{p}{(}\DUrole{sc}{'b'}\DUrole{p}{,}~\\
~~~~~~~~~~~~~\DUrole{n}{box}~\DUrole{n}{Nil}\DUrole{p}{));}~\\
~~~\DUrole{n}{println}\DUrole{o}{!}\DUrole{p}{(}\DUrole{s}{\textquotedbl{}\{\}\textquotedbl{}}\DUrole{p}{,}~\DUrole{n}{len}\DUrole{p}{(}\DUrole{n}{l}\DUrole{p}{));}~\\
\DUrole{p}{\}}
}
 & 
{\ttfamily \raggedright \noindent
\DUrole{c}{(*~OCaml~*)}~\\
\DUrole{k}{type}~\DUrole{k}{'}\DUrole{n}{t}~\DUrole{n}{lst}~\DUrole{o}{=}~\\
~~~~\DUrole{nc}{Nil}~\\
~~\DUrole{o}{|}~\DUrole{nc}{Cons}~\DUrole{k}{of}~\DUrole{k}{'}\DUrole{n}{t}~\DUrole{o}{*}~\DUrole{k}{'}\DUrole{n}{t}~\DUrole{n}{lst}~\\
~\\
\DUrole{k}{let}~\DUrole{n}{len}~\DUrole{n}{l}~\DUrole{o}{=}~\\
~~\DUrole{k}{match}~\DUrole{n}{l}~\DUrole{k}{with}~\\
~~\DUrole{o}{|}~\DUrole{nc}{Nil}~\DUrole{o}{->}~\DUrole{mi}{0}~\\
~~\DUrole{o}{|}~\DUrole{nc}{Cons}\DUrole{o}{(\_,}~\DUrole{n}{x}\DUrole{o}{)}~\DUrole{o}{->}~\DUrole{mi}{1}~\DUrole{o}{+}~\DUrole{n}{len}~\DUrole{n}{x}~\\
~\\
~\\
~\\
~\\
\DUrole{k}{let}~\DUrole{n}{l}~\DUrole{o}{=}~\DUrole{nc}{Cons}\DUrole{o}{(}\DUrole{sc}{'a'}\DUrole{o}{,}~\\
~~~~~~~~~~\DUrole{nc}{Cons}\DUrole{o}{(}\DUrole{sc}{'b'}\DUrole{o}{,}~\\
~~~~~~~~~~~~~~\DUrole{nc}{Nil}\DUrole{o}{));}~\\
\DUrole{n}{print\_int}~\DUrole{o}{(}\DUrole{n}{len}~\DUrole{n}{l}\DUrole{o}{)}
}
 & 
{\ttfamily \raggedright \noindent
\DUrole{cm}{\{-~Haskell~-\}}~\\
\DUrole{kr}{data}~\DUrole{kt}{Lst}~\DUrole{n}{t}~\DUrole{ow}{=}~\\
~~~~\DUrole{kt}{Nil}~\\
~~\DUrole{o}{|}~\DUrole{kt}{Cons}~\DUrole{n}{t}~\DUrole{p}{(}\DUrole{kt}{Lst}~\DUrole{n}{t}\DUrole{p}{)}~\\
~\\
\DUrole{kr}{let}~\DUrole{n}{len}~\DUrole{n}{l}~\DUrole{ow}{=}~\\
~~\DUrole{kr}{case}~\DUrole{n}{l}~\DUrole{kr}{of}~\\
~~~~\DUrole{kt}{Nil}~\DUrole{ow}{->}~\DUrole{mi}{0}~\\
~~~~\DUrole{kt}{Cons}~\DUrole{kr}{\_}~\DUrole{n}{x}~\DUrole{ow}{->}~\DUrole{mi}{1}~\DUrole{o}{+}~\DUrole{n}{len}~\DUrole{n}{x}~\\
~\\
~\\
~\\
\DUrole{nf}{main}~\DUrole{ow}{=}~\DUrole{kr}{do}~\\
~~~~~~~~~\DUrole{kr}{let}~\DUrole{n}{l}~\DUrole{ow}{=}~\DUrole{kt}{Cons}~\DUrole{n}{'a'}~\\
~~~~~~~~~~~~~~~~~~~~~~\DUrole{p}{(}\DUrole{kt}{Cons}~\DUrole{n}{'b'}~\\
~~~~~~~~~~~~~~~~~~~~~~~~~~~~\DUrole{kt}{Nil}\DUrole{p}{)}~\\
~~~~~~~~~\DUrole{n}{putStrLn}~\DUrole{o}{\$}~\DUrole{n}{show}~\DUrole{n}{l}
}
 \\
\end{longtable*}

\section{Shared objects: Rc and Arc%
  \label{shared-objects-rc-and-arc}%
}

In complement to the facilities offered by \texttt{box}, the Rust standard
library also implements two implementations of \textbf{reference counted
wrappers to immutable objects} that can be referred to from multiple
owners:
\begin{quote}{\ttfamily \raggedright \noindent
\DUrole{k}{use}~\DUrole{n}{std}\DUrole{o}{::}\DUrole{n}{rc}\DUrole{o}{::}\DUrole{n}{Rc}\DUrole{p}{;}~\\
\DUrole{k}{fn}~\DUrole{n}{main}\DUrole{p}{()}~\DUrole{p}{\{}~\\
~~~~\DUrole{k}{let}~\DUrole{n}{b}~\DUrole{o}{=}~\DUrole{n}{Rc}\DUrole{o}{::}\DUrole{n}{new}\DUrole{p}{(}\DUrole{m}{3}\DUrole{k}{i}\DUrole{p}{);}~\\
~~~~\DUrole{k}{let}~\DUrole{n}{c}~\DUrole{o}{=}~\DUrole{o}{\&}\DUrole{n}{b}\DUrole{p}{;}~\\
~~~~\DUrole{n}{println}\DUrole{p}{(}\DUrole{s}{\textquotedbl{}\{\}~\{\}\textquotedbl{}}\DUrole{p}{,}~\DUrole{n}{b}\DUrole{p}{,}~\DUrole{n}{c}\DUrole{p}{);}~\DUrole{c1}{//~OK\\
}\DUrole{p}{\}}
}
\end{quote}

The use of reference counts ensures that an object gets deallocated
exactly when its last reference is dropped.  The trade-off with boxes
is that the reference counter must be updated every time a new
reference is created or a reference is dropped.

Two implementations are provided: \texttt{std::rc::Rc} and
\texttt{std::arc::Arc}. Both offer the same interface. The reason for this
duplication is to offer a controllable trade-off over performance to
programmers: \texttt{Rc} does not use memory atomics, so it is more
lightweight and thus faster, however it cannot be shared across
threads. \texttt{Arc} does use atomics, is thus slightly less efficient
than \texttt{Rc}, but can be used to share data across threads.

\section{Macros and meta-programming%
  \label{macros-and-meta-programming}%
}

The basic syntax of a macro definition is as follows:
\begin{quote}{\ttfamily \raggedright \noindent
\DUrole{n}{macro\_rules}\DUrole{o}{!}~\DUrole{n}{MACRONAME}~\DUrole{p}{(}~\\
~~~~\DUrole{p}{(}\DUrole{n}{PATTERN}\DUrole{p}{...)}~\DUrole{o}{=>}~\DUrole{p}{(}\DUrole{n}{EXPANSION}\DUrole{p}{...);}~\\
~~~~\DUrole{p}{(}\DUrole{n}{PATTERN}\DUrole{p}{...)}~\DUrole{o}{=>}~\DUrole{p}{(}\DUrole{n}{EXPANSION}\DUrole{p}{...);}~\\
~~~~\DUrole{p}{...}~\\
\DUrole{p}{)}
}
\end{quote}

For example, the following macro defines a Pascal-like \texttt{for} loop:
\begin{quote}{\ttfamily \raggedright \noindent
\DUrole{n}{macro\_rules}\DUrole{o}{!}~\DUrole{n}{pfor}~\DUrole{p}{(}~\\
~~~\DUrole{p}{(}\DUrole{n}{\$x}\DUrole{o}{:}\DUrole{n}{ident}~\DUrole{o}{=}~\DUrole{n}{\$s}\DUrole{o}{:}\DUrole{n}{expr}~\DUrole{n}{to}~\DUrole{n}{\$e}\DUrole{o}{:}\DUrole{n}{expr}~\DUrole{n}{\$body}\DUrole{o}{:}\DUrole{n}{expr}\DUrole{p}{)}~\\
~~~~~~~\DUrole{o}{=>}~\DUrole{p}{(}\DUrole{k}{match}~\DUrole{n}{\$e}~\DUrole{p}{\{}~\DUrole{n}{e}~\DUrole{o}{=>}~\DUrole{p}{\{}~\\
~~~~~~~~~~~~~~\DUrole{k}{let}~\DUrole{k}{mut}~\DUrole{n}{\$x}~\DUrole{o}{=}~\DUrole{n}{\$s}\DUrole{p}{;}~\\
~~~~~~~~~~~~~~\DUrole{k}{loop}~\DUrole{p}{\{}~\\
~~~~~~~~~~~~~~~~~~~\DUrole{n}{\$body}\DUrole{p}{;}~\\
~~~~~~~~~~~~~~~~~~~\DUrole{n}{\$x}~\DUrole{o}{+=}~\DUrole{m}{1}\DUrole{k}{}\DUrole{p}{;}~\\
~~~~~~~~~~~~~~~~~~~\DUrole{k}{if}~\DUrole{n}{\$x}~\DUrole{o}{>}~\DUrole{n}{e}~\DUrole{p}{\{}~\DUrole{k}{break}\DUrole{p}{;}~\DUrole{p}{\}}~\\
~~~~~~~~~~~~~~\DUrole{p}{\}}~\\
~~~~~~~~~~\DUrole{p}{\}\});}~\\
\DUrole{p}{);}~\\
~\\
\DUrole{c1}{//~Example~use:\\
}\DUrole{k}{fn}~\DUrole{n}{main}\DUrole{p}{()}~\DUrole{p}{\{}~\\
~~~~~\DUrole{n}{pfor}\DUrole{o}{!}\DUrole{p}{(}\DUrole{n}{i}~\DUrole{o}{=}~\DUrole{m}{0}\DUrole{k}{}~\DUrole{n}{to}~\DUrole{m}{10}\DUrole{k}{}~\DUrole{p}{\{}~\\
~~~~~~~~~\DUrole{n}{println}\DUrole{o}{!}\DUrole{p}{(}\DUrole{s}{\textquotedbl{}\{\}\textquotedbl{}}\DUrole{p}{,}~\DUrole{n}{i}\DUrole{p}{);}~\\
~~~~~\DUrole{p}{\});}~\\
\DUrole{p}{\}}
}
\end{quote}

Note how this macro uses the \texttt{match} statement to assign a local
name to an expression, so that it does not get evaluated more than once.

Like in Scheme, macros can be recursive. For example, the following
macro uses recursion to implement \texttt{pfor} both with and without \texttt{step}:
\begin{quote}{\ttfamily \raggedright \noindent
\DUrole{n}{macro\_rules}\DUrole{o}{!}~\DUrole{n}{pfor}~\DUrole{p}{(}~\\
~~~\DUrole{p}{(}\DUrole{n}{\$x}\DUrole{o}{:}\DUrole{n}{ident}~\DUrole{o}{=}~\DUrole{n}{\$s}\DUrole{o}{:}\DUrole{n}{expr}~\DUrole{n}{to}~\DUrole{n}{\$e}\DUrole{o}{:}\DUrole{n}{expr}~\DUrole{n}{step}~\DUrole{n}{\$st}\DUrole{o}{:}\DUrole{n}{expr}~\DUrole{n}{\$body}\DUrole{o}{:}\DUrole{n}{expr}\DUrole{p}{)}~\\
~~~~~~~\DUrole{o}{=>}~\DUrole{p}{(}\DUrole{k}{match}~\DUrole{n}{\$e}\DUrole{p}{,}\DUrole{n}{\$st}~\DUrole{p}{\{}~\DUrole{n}{e}\DUrole{p}{,}~\DUrole{n}{st}~\DUrole{o}{=>}~\DUrole{p}{\{}~\\
~~~~~~~~~~~~~~\DUrole{k}{let}~\DUrole{k}{mut}~\DUrole{n}{\$x}~\DUrole{o}{=}~\DUrole{n}{\$s}\DUrole{p}{;}~\\
~~~~~~~~~~~~~~\DUrole{k}{loop}~\DUrole{p}{\{}~\\
~~~~~~~~~~~~~~~~~~~\DUrole{n}{\$body}\DUrole{p}{;}~\\
~~~~~~~~~~~~~~~~~~~\DUrole{n}{\$x}~\DUrole{o}{+=}~\DUrole{n}{st}\DUrole{p}{;}~\\
~~~~~~~~~~~~~~~~~~~\DUrole{k}{if}~\DUrole{n}{\$x}~\DUrole{o}{>}~\DUrole{n}{e}~\DUrole{p}{\{}~\DUrole{k}{break}\DUrole{p}{;}~\DUrole{p}{\}}~\\
~~~~~~~~~~~~~~\DUrole{p}{\}}~\\
~~~~~~~~~~\DUrole{p}{\}\});}~\\
~~~\DUrole{p}{(}\DUrole{n}{\$x}\DUrole{o}{:}\DUrole{n}{ident}~\DUrole{o}{=}~\DUrole{n}{\$s}\DUrole{o}{:}\DUrole{n}{expr}~\DUrole{n}{to}~\DUrole{n}{\$e}\DUrole{o}{:}\DUrole{n}{expr}~\DUrole{n}{\$body}\DUrole{o}{:}\DUrole{n}{expr}\DUrole{p}{)}~\\
~~~~~~\DUrole{o}{=>}~\DUrole{p}{(}\DUrole{n}{pfor}\DUrole{o}{!}\DUrole{p}{(}\DUrole{n}{\$x}~\DUrole{o}{=}~\DUrole{n}{\$s}~\DUrole{n}{to}~\DUrole{n}{\$e}~\DUrole{n}{step}~\DUrole{m}{1}\DUrole{k}{}~\DUrole{n}{\$body}\DUrole{p}{));}~\\
\DUrole{p}{);}~\\
~\\
\DUrole{c1}{//~Example~use:\\
}\DUrole{k}{fn}~\DUrole{n}{main}\DUrole{p}{()}~\DUrole{p}{\{}~\\
~~~~~\DUrole{n}{pfor}\DUrole{o}{!}\DUrole{p}{(}\DUrole{n}{i}~\DUrole{o}{=}~\DUrole{m}{0}\DUrole{k}{}~\DUrole{n}{to}~\DUrole{m}{10}\DUrole{k}{}~\DUrole{n}{step}~\DUrole{m}{2}\DUrole{k}{}~\DUrole{p}{\{}~\\
~~~~~~~~~\DUrole{n}{println}\DUrole{o}{!}\DUrole{p}{(}\DUrole{s}{\textquotedbl{}\{\}\textquotedbl{}}\DUrole{p}{,}~\DUrole{n}{i}\DUrole{p}{);}~\\
~~~~~\DUrole{p}{\});}~\\
\DUrole{p}{\}}
}
\end{quote}

Macros can also be variadic, in that arbitrary repetitions of a syntax
form can be captured by one macro argument. For example, the following
macro invokes \texttt{println!} on each of its arguments, which can
be of arbitrary type:
\begin{quote}{\ttfamily \raggedright \noindent
\DUrole{n}{macro\_rules}\DUrole{o}{!}~\DUrole{n}{printall}~\DUrole{p}{(}~\\
~~~~~\DUrole{p}{(}~\DUrole{n}{\$}\DUrole{p}{(}~\DUrole{n}{\$arg}\DUrole{o}{:}\DUrole{n}{expr}~\DUrole{p}{),}\DUrole{o}{*}~~\DUrole{p}{)}~\DUrole{o}{=>}~\DUrole{p}{(}~\\
~~~~~~~~~~\DUrole{n}{\$}\DUrole{p}{(}~\DUrole{n}{println}\DUrole{o}{!}\DUrole{p}{(}\DUrole{s}{\textquotedbl{}\{\}\textquotedbl{}}\DUrole{p}{,}~\DUrole{n}{\$arg}\DUrole{p}{)}~\DUrole{p}{);}\DUrole{o}{*}~\\
~~~~~\DUrole{p}{);}~\\
\DUrole{p}{);}~\\
~\\
\DUrole{c1}{//~example~use:\\
}\DUrole{k}{fn}~\DUrole{n}{main}\DUrole{p}{()}~\DUrole{p}{\{}~\\
~~~~\DUrole{n}{printall}\DUrole{o}{!}\DUrole{p}{(}\DUrole{s}{\textquotedbl{}hello\textquotedbl{}}\DUrole{p}{,}~\DUrole{m}{42}\DUrole{k}{}\DUrole{p}{,}~\DUrole{m}{3.14}\DUrole{k}{}\DUrole{p}{);}~\\
\DUrole{p}{\}}
}
\end{quote}

The syntax works as follows: on the left hand side (pattern) the form
\texttt{\$( PAT )DELIM*} matches zero or more occurrences of PAT delimited
by \texttt{DELIM}; on the right hand side (expansion), the form \texttt{\$( TEXT
)DELIM*} expands to one or more repetitions of \texttt{TEXT} separated by
\texttt{DELIM}. The number of repetitions of the expansion is determined by
the number of matches of the enclosed macro argument(s). In the
example, each argument (separated by commas) is substituted by a
corresponding invocation of \texttt{println!}, separated by semicolons.

\section{Literals%
  \label{literals}%
}

Rust provides various lexical forms for number literals:

\setlength{\DUtablewidth}{\linewidth}
\begin{longtable*}[c]{p{0.193\DUtablewidth}p{0.138\DUtablewidth}p{0.116\DUtablewidth}p{0.271\DUtablewidth}p{0.237\DUtablewidth}}
\textbf{%
Rust syntax
} & \textbf{%
Same as
} & \textbf{%
Type
} & \textbf{%
Haskell equivalent
} & \textbf{%
OCaml equivalent
} \\
\endfirsthead
\textbf{%
Rust syntax
} & \textbf{%
Same as
} & \textbf{%
Type
} & \textbf{%
Haskell equivalent
} & \textbf{%
OCaml equivalent
} \\
\endhead
\multicolumn{5}{c}{\hfill ... continued on next page} \\
\endfoot
\endlastfoot

\texttt{123i}
 &  & 
int
 & 
\texttt{123 :: Int}
 & 
\texttt{123}
 \\

\texttt{123u}
 &  & 
uint
 & 
\texttt{123 :: Word}
 &  \\

\texttt{123i8}
 &  & 
i8
 & 
\texttt{123 :: Int8}
 &  \\

\texttt{123u8}
 &  & 
u8
 & 
\texttt{123 :: Word8}
 & 
\texttt{Char.chr 123}
 \\

\texttt{123i16}
 &  & 
i16
 & 
\texttt{123 :: Int16}
 &  \\

\texttt{123u16}
 &  & 
u16
 & 
\texttt{123 :: Word16}
 &  \\

\texttt{123i32}
 &  & 
i32
 & 
\texttt{123 :: Int32}
 & 
\texttt{123l}
 \\

\texttt{123u32}
 &  & 
u32
 & 
\texttt{123 :: Word32}
 &  \\

\texttt{123i64}
 &  & 
i64
 & 
\texttt{123 :: Int64}
 & 
\texttt{123L}
 \\

\texttt{123u64}
 &  & 
u64
 & 
\texttt{123 :: Word64}
 &  \\

\texttt{1\_2\_3\_4}
 & 
\texttt{1234}
 & 
(integer)
 &  &  \\

\texttt{1234\_i}
 & 
\texttt{1234i}
 & 
int
 &  &  \\

\texttt{0x1234}
 & 
\texttt{4660}
 & 
(integer)
 & 
\texttt{0x1234}
 & 
\texttt{0x1234}
 \\

\texttt{0x1234u16}
 & 
\texttt{4660u16}
 & 
u16
 & 
\texttt{0x1234 :: Word16}
 &  \\

\texttt{0b1010}
 & 
\texttt{10}
 & 
(integer)
 &  & 
\texttt{0b1010}
 \\

\texttt{0o1234}
 & 
\texttt{668}
 & 
(integer)
 & 
\texttt{0o1234}
 & 
\texttt{0o1234}
 \\

\texttt{b'a'}
 & 
\texttt{97u8}
 & 
u8
 &  & 
\texttt{'a'}
 \\

\texttt{b\textquotedbl{}a\textquotedbl{}}
 & 
\texttt{{[}97u8{]}}
 & 
{[}u8{]}
 &  &  \\

\texttt{12.34}
 &  & 
(float)
 & 
\texttt{12.34}
 &  \\

\texttt{12.34f32}
 &  & 
f32
 & 
\texttt{12.34 :: Float}
 &  \\

\texttt{12.34f64}
 &  & 
f64
 & 
\texttt{12.34 :: Double}
 & 
\texttt{12.34}
 \\

\texttt{12e34}
 & 
\texttt{1.2e35}
 & 
(float)
 & 
\texttt{12e34}
 &  \\

\texttt{12E34}
 & 
\texttt{1.2e35}
 & 
(float)
 & 
\texttt{12E34}
 &  \\

\texttt{12E+34}
 & 
\texttt{1.2e35}
 & 
(float)
 & 
\texttt{12E+34}
 &  \\

\texttt{12E-34}
 & 
\texttt{1.2e-33}
 & 
(float)
 & 
\texttt{12E-34}
 &  \\

\texttt{1\_2e34}
 & 
\texttt{12e34}
 & 
(float)
 &  &  \\

\texttt{1\_2e3\_4}
 & 
\texttt{12e34}
 & 
(float)
 &  &  \\
\end{longtable*}

Escapes in character, byte and string literals:

\setlength{\DUtablewidth}{\linewidth}
\begin{longtable*}[c]{p{0.203\DUtablewidth}p{0.145\DUtablewidth}p{0.145\DUtablewidth}p{0.110\DUtablewidth}p{0.203\DUtablewidth}p{0.145\DUtablewidth}}
\textbf{%
Syntax
} & \textbf{%
Same as
} & \textbf{%
Syntax
} & \textbf{%
Same as
} & \textbf{%
Syntax
} & \textbf{%
Same as
} \\
\endfirsthead
\textbf{%
Syntax
} & \textbf{%
Same as
} & \textbf{%
Syntax
} & \textbf{%
Same as
} & \textbf{%
Syntax
} & \textbf{%
Same as
} \\
\endhead
\multicolumn{6}{c}{\hfill ... continued on next page} \\
\endfoot
\endlastfoot

\texttt{'\textbackslash{}x61'}
 & 
\texttt{'a'}
 & 
\texttt{b'\textbackslash{}x61'}
 & 
97u8
 & 
\texttt{\textquotedbl{}\textbackslash{}x61\textquotedbl{}}
 & 
\texttt{\textquotedbl{}a\textquotedbl{}}
 \\

\texttt{'\textbackslash{}\textbackslash{}'}
 & 
\texttt{'\textbackslash{}x5c'}
 & 
\texttt{b'\textbackslash{}\textbackslash{}'}
 & 
92u8
 & 
\texttt{\textquotedbl{}\textbackslash{}\textbackslash{}\textquotedbl{}}
 & 
\texttt{\textquotedbl{}\textbackslash{}x5c\textquotedbl{}}
 \\

\texttt{'\textbackslash{}'{}'}
 & 
\texttt{'\textbackslash{}x27'}
 & 
\texttt{b'\textbackslash{}'{}'}
 & 
39u8
 & 
\texttt{\textquotedbl{}\textbackslash{}\textquotedbl{}\textquotedbl{}}
 & 
\texttt{\textquotedbl{}\textbackslash{}x22\textquotedbl{}}
 \\

\texttt{'\textbackslash{}0'}
 & 
\texttt{'\textbackslash{}x00'}
 & 
\texttt{b'\textbackslash{}0'}
 & 
0u8
 & 
\texttt{\textquotedbl{}\textbackslash{}0\textquotedbl{}}
 & 
\texttt{\textquotedbl{}\textbackslash{}x00\textquotedbl{}}
 \\

\texttt{'\textbackslash{}t'}
 & 
\texttt{'\textbackslash{}x09'}
 & 
\texttt{b'\textbackslash{}t'}
 & 
9u8
 & 
\texttt{\textquotedbl{}\textbackslash{}t\textquotedbl{}}
 & 
\texttt{\textquotedbl{}\textbackslash{}x09\textquotedbl{}}
 \\

\texttt{'\textbackslash{}n'}
 & 
\texttt{'\textbackslash{}x0a'}
 & 
\texttt{b'\textbackslash{}n'}
 & 
10u8
 & 
\texttt{\textquotedbl{}\textbackslash{}n\textquotedbl{}}
 & 
\texttt{\textquotedbl{}\textbackslash{}x0a\textquotedbl{}}
 \\

\texttt{'\textbackslash{}r'}
 & 
\texttt{'\textbackslash{}x0d'}
 & 
\texttt{b'\textbackslash{}r'}
 & 
13u8
 & 
\texttt{\textquotedbl{}\textbackslash{}r\textquotedbl{}}
 & 
\texttt{\textquotedbl{}\textbackslash{}x0d\textquotedbl{}}
 \\

\texttt{'\textbackslash{}u0123'}
 &  &  &  & 
\texttt{\textquotedbl{}\textbackslash{}u0123\textquotedbl{}}
 &  \\

\texttt{'\textbackslash{}U00012345'}
 &  &  &  & 
\texttt{\textquotedbl{}\textbackslash{}U00012345\textquotedbl{}}
 &  \\
\end{longtable*}

Note that the other common escapes in the C family (\texttt{\textbackslash{}a}, \texttt{\textbackslash{}f},
etc.) are not valid in Rust, nor are octal escapes (eg. \texttt{\textbackslash{}0123}).

Finally, Rust like Python supports raw strings and multiple string delimiters, to
avoid quoting occurrences of the delimiters within the string:

\setlength{\DUtablewidth}{\linewidth}
\begin{longtable*}[c]{p{0.202\DUtablewidth}p{0.146\DUtablewidth}p{0.230\DUtablewidth}p{0.371\DUtablewidth}}
\textbf{%
Syntax
} & \textbf{%
String value
} & \textbf{%
Syntax
} & \textbf{%
Value
} \\
\endfirsthead
\textbf{%
Syntax
} & \textbf{%
String value
} & \textbf{%
Syntax
} & \textbf{%
Value
} \\
\endhead
\multicolumn{4}{c}{\hfill ... continued on next page} \\
\endfoot
\endlastfoot

\texttt{\textquotedbl{}foo\textquotedbl{}}
 & 
\texttt{foo}
 & 
\texttt{b\textquotedbl{}foo\textquotedbl{}}
 & 
\texttt{{[}102u8,111,111{]}}
 \\

\texttt{\textquotedbl{}fo\textbackslash{}\textquotedbl{}o\textquotedbl{}}
 & 
\texttt{fo\textquotedbl{}o}
 & 
\texttt{b\textquotedbl{}fo\textbackslash{}\textquotedbl{}o\textquotedbl{}}
 & 
\texttt{{[}102u8,111,34,111{]}}
 \\

\texttt{r\textquotedbl{}fo\textbackslash{}n\textquotedbl{}}
 & 
\texttt{fo\textbackslash{}n}
 & 
\texttt{rb\textquotedbl{}fo\textbackslash{}n\textquotedbl{}}
 & 
\texttt{{[}102u8,111,92,110{]}}
 \\

\texttt{r\#\textquotedbl{}fo\textbackslash{}\textquotedbl{}o\textquotedbl{}\#}
 & 
\texttt{fo\textbackslash{}\textquotedbl{}o}
 & 
\texttt{rb\#\textquotedbl{}fo\textbackslash{}\textquotedbl{}o\textquotedbl{}\#}
 & 
\texttt{{[}102u8,111,92,34,111{]}}
 \\

\texttt{\textquotedbl{}foo\#\textbackslash{}\textquotedbl{}\#bar\textquotedbl{}}
 & 
\texttt{foo\#\textquotedbl{}\#bar}
 & 
\texttt{b\textquotedbl{}foo\#\textbackslash{}\textquotedbl{}\#bar\textquotedbl{}}
 & 
\texttt{{[}102u8,111,111,35,34,35,98,97,114{]}}
 \\

\texttt{r\#\#\textquotedbl{}foo\#\textquotedbl{}\#bar\textquotedbl{}\#\#}
 & 
\texttt{foo\#\textquotedbl{}\#bar}
 & 
\texttt{rb\#\#\textquotedbl{}foo\#\textquotedbl{}\#bar\textquotedbl{}\#\#}
 & 
\texttt{{[}102u8,111,111,35,34,35,98,97,114{]}}
 \\
\end{longtable*}

\section{Acknowledgements%
  \label{acknowledgements}%
}

Many thanks to the numerous commenters on Reddit and Hacker news who
provided high-quality comments and substantively contributed to
improving upon the first version of this article.

\section{References%
  \label{references}%
}
\begin{itemize}

\item \href{http://doc.rust-lang.org/0.10/tutorial.html}{The Rust Language Tutorial}, version 0.10, April 2014.

\item \href{http://doc.rust-lang.org/0.11.0/tutorial.html}{The Rust Language Tutorial}, version 0.11.0, July 2014.

\item \href{http://doc.rust-lang.org/0.11.0/guide.html}{The Rust Guide}, version 0.11.0, July 2014.

\item Aaron Turon, \href{https://aturon.github.io/}{Rust Guidelines}, 2014.

\item Will Yager. \href{http://yager.io/programming/go.html}{Why Go is not good}, 2014. Explains how Go lacks many
features found in Rust and thereby fails to be a modern functional
language.

\item Edward Z. Yang, \href{http://blog.ezyang.com/2010/10/ocaml-for-haskellers/}{OCaml for Haskellers}, October 2010.

\item Raphael ‘kena’ Poss, \href{http://science.raphael.poss.name/haskell-for-ocaml-programmers.html}{Haskell for OCaml programmers}, March 2014.

\item Xavier Leroy et al., \href{http://caml.inria.fr/pub/docs/manual-ocaml-4.01/}{The OCaml system release 4.01}, September 2013.

\end{itemize}

\DUtransition

\section{Copyright and licensing%
  \label{copyright-and-licensing}%
}

Copyright © 2014, Raphael ‘kena’ Poss.
Permission is granted to distribute, reuse and modify this document
according to the terms of the Creative Commons Attribution-ShareAlike
4.0 International License.  To view a copy of this license, visit
\url{http://creativecommons.org/licenses/by-sa/4.0/}.

\DUtransition

\href{http://www.structured-commons.org}{SC} fingerprint: \texttt{fp:khrVQc-ggfInTTQIS6\_5KlHvvUNOVjiyNXkNPmW3kr6ydw}

\end{document}